\documentclass[aps,floatfix,prd,twocolumn,letterpaper,eqsecnum,amsmath,amssymb,nofootinbib,superscriptaddress,longbibliography]{revtex4-2}
\usepackage{amsmath,amssymb,mathrsfs,graphicx,booktabs}
\usepackage[colorlinks=true,linkcolor=blue,citecolor=blue,urlcolor=blue]{hyperref}
\newcommand{\mainEq}[2]{\eqref{#1}}

\begin{document}
\title{Thermodynamic Degeneracy and Distinct Horizon Geometries of Rotating Kaluza–Klein Black Holes in Independent External Electric and Magnetic Fields}
\author{Hai Huang}
\email{haihuang@gyu.edu.cn}
\affiliation{College of Mechanical Engineering, Guiyang University,
103 Jianlong Rd., Guiyang, Guizhou 550005, China}

\author{Juhua Chen}
\email{jhchen@hunnu.edu.cn}
\affiliation{Department of Physics, Hunan Normal University,
36 Lushan Rd., Changsha, Hunan 410081, China}
\begin{abstract}
We construct an exact finite-rotation solution of
four-dimensional Einstein--Maxwell--dilaton theory
at the Kaluza--Klein coupling, immersed
in independent external electric and magnetic fields. Exact algebraic balance
identities reveal two distinct regular nonextremal
branches with identical horizon area, quasi-local
angular momentum, and electromagnetic fluxes. Under
a common, explicitly specified reference prescription
for the axial clock, rotational frame, and electromagnetic
potentials, the two branches also have identical
thermodynamic readouts, including energy, temperature,
angular velocity, and electric and magnetic potentials.
Nevertheless, the two branches possess nonisometric
intrinsic horizon geometries, with every positive
axisymmetric scalar-Laplacian eigenvalue strictly
ordered between them. Thus, identical thermodynamic
readouts do not uniquely determine the intrinsic
geometry of a black-hole horizon.
We further establish that the exterior admits consistent
asymptotic matching to two distinct exact rotating
cylindrical backgrounds associated with its northern
and southern polar regions. Their metric, dilaton, and
electromagnetic fields become compatible in the angular
overlap region, with uniform matching extending to both
symmetry axes. Every regular noncritical exterior,
including the supercritical branches, is globally
hyperbolic.
\end{abstract}
\maketitle

\section{Introduction}

Can a black hole's complete thermodynamic readouts
uniquely determine its intrinsic horizon geometry?
Entropy fixes the total horizon area, while angular
momentum and electromagnetic charges provide global
information~\cite{BardeenCarterHawking1973,Wald1993,
AshtekarKrishnan2002}. The intrinsic horizon metric,
however, also encodes the local distribution of
curvature, orbit lengths, and the spectrum of geometric
operators. Whether these additional data are fixed by
black-hole thermodynamics is a fundamental question
concerning the relation between horizon geometry and
thermodynamic observables.

Previous studies have established that external fields
and surrounding matter can deform black-hole horizons
\cite{Yazadjiev2001,Yazadjiev2006,Yazadjiev2013Charged,
Yazadjiev2013,GibbonsMujtabaPope2013,
GibbonsPangPope2014,Shoom2015}. The geometry of
isolated horizons provides a complementary description:
invariant mass and angular-momentum multipoles
characterize the intrinsic geometry beyond its lowest
global moments
\cite{AshtekarBeetleLewandowski2002,
AshtekarEnglePawlowskiVanDenBroeck2004}.
Under appropriate conditions, the complete multipole
data determine the intrinsic horizon geometry, whereas
a finite collection of integrated thermodynamic
observables need not contain enough information to
reconstruct the full angular geometry. A related
geometric perspective comes from spectral theory:
the eigenvalues of intrinsic differential operators
provide further invariant information about a horizon's
geometry~\cite{AbreuFreitas2002}.
These developments motivate a sharper question:
can two regular rotating black holes possess identical
complete thermodynamic readouts while exhibiting
strictly different horizon geometries and geometric
spectra? Here, thermodynamic degeneracy is understood relative
to a specified common reference prescription: distinct
regular black-hole solutions share identical
thermodynamic readouts.

A natural setting for investigating this question is
provided by rotating black holes in
Einstein--Maxwell--dilaton (EMD) theory
\cite{GibbonsMaeda1988,GarfinkleHorowitzStrominger1991,
HorneHorowitz1992,Rasheed1995,Larsen2000,
MatosMora1997,KunzMaisonNavarroLeridaViebahn2006,
Yazadjiev2001,Yazadjiev2006,Yazadjiev2013Charged,
Yazadjiev2013,HerdeiroRaduCostaFilho2025,
TomizawaSakamotoSuzuki2025,TomizawaEtAl2026}.
At the Kaluza--Klein (KK) coupling $\alpha^2=3$,
the stationary axisymmetric reduction possesses an
enhanced $SL(3,\mathbb{R})$ symmetry
\cite{Maison1979,GGK,Galtsov1995}.
Rotating black holes in external electromagnetic
fields and magnetized solutions have also been
constructed in Einstein--Maxwell and dilatonic
theories
\cite{Wald1974,Melvin1964,Harrison1968,
Ernst1968Vacuum,Ernst1968Electrovac,Ernst1976,
ErnstWild1976,GibbonsMujtabaPope2013,
GibbonsPangPope2014,PodolskyOvcharenko2025,
OvcharenkoPodolsky2026,DiPintoKlemmVigano2025}.
Static EMD configurations with independent external
electric and magnetic fields have recently been
studied~\cite{MaWuLu2026}. The finite-rotation
problem with independent external electric and
magnetic fields in a single Maxwell sector therefore
provides a distinct setting for investigating the
relation between thermodynamic observables and
horizon geometry.

In this Letter, we address this question by constructing
an exact finite-rotation solution of four-dimensional
EMD theory at the KK coupling, immersed in independent
external electric and magnetic fields. An exact
algebraic reconstruction reveals two distinct regular
branches with the same horizon area, quasi-local
angular momentum, and electromagnetic fluxes.
Under a common, explicitly specified thermodynamic
reference prescription, the two branches also possess
identical energy, temperature, angular velocity, and
electric and magnetic potentials. The energy is the
Hamiltonian associated with an integrable surface-charge
variation on the specified finite-dimensional family
of exact solutions. Nevertheless, we prove that their
intrinsic horizon geometries are strictly different.
Moreover, the corresponding positive axisymmetric
scalar-Laplacian eigenvalues are strictly ordered
between the two branches.

In addition, the exact exterior admits two distinct
rotating cylindrical backgrounds associated with its
northern and southern polar regions. We show that
these backgrounds are independently exact solutions
of the four-dimensional EMD equations and that their
metric, dilaton, and electromagnetic fields become
asymptotically compatible in their angular overlap
region. Uniform matching holds up to both symmetry
axes, providing two consistent asymptotic descriptions
of the same connected spatial end. We also establish
the global causal structure of the regular exterior,
including the identification of its future outer
horizon relative to a specified axial future escape end.

\section{Exact rotating double-field solution}

We consider EMD theory with action
\begin{equation}
I=\frac{1}{16\pi}\int d^4x\sqrt{-g}
\left[R-2(\nabla\phi)^2-e^{-2\sqrt3\phi}F^2\right].
\label{eq:action}
\end{equation}
At the KK coupling, the rotating solution with independent external electric and magnetic fields admits the compact form
\begin{equation}
\begin{aligned}
ds^2={}&H\left[
-\frac{\Sigma\Delta}{\mathscr A}dt^2
+\Sigma\left(\frac{d\rho^2}{\Delta}
+d\theta^2\right)\right]\\
&+\frac{s}{H}(d\varphi-\omega dt)^2.
\end{aligned}
\label{eq:metric}
\end{equation}
The metric functions are
\begin{equation}
\begin{aligned}
\Sigma&=\rho^2+a^2\cos^2\theta,\: \qquad \Delta=\rho^2-2m\rho+a^2,\\
\mathscr A&=(\rho^2+a^2)^2-a^2\Delta\sin^2\theta,\quad s=\frac{\mathscr A\sin^2\theta}{\Sigma},\\
\mathcal{W}&=ma\cos\theta(3-\cos^2\theta)+\frac{ma^3\cos\theta\sin^4\theta}{\Sigma}.
\end{aligned}
\label{eq:metric-functions}
\end{equation}
and
\begin{equation}
\begin{aligned}
H_+&=(1+E_0B_0\mathcal{W})^2+E_0^2s
+\frac{E_0^2B_0^2s^2}{4},\\
H_-&=(1-E_0B_0\mathcal{W})^2+B_0^2s
+\frac{E_0^2B_0^2s^2}{4},\\
H&=\sqrt{H_+H_-}.
\end{aligned}
\label{eq:field-factors}
\end{equation}
Here $\omega$ is the frame-dragging function, the mass parameter \(m\) and rotation parameter \(a\) satisfy \(m>|a|\). Because the exterior is not asymptotically flat, \(m\) is not an ADM mass. The independent amplitudes \(E_0\) and \(B_0\) specify the external electric and magnetic fields of a single Maxwell field. 

The associated dilaton, Maxwell potential and frame-dragging function are

\begin{align}
\phi&=\frac{\sqrt3}{4}\log\frac{H_+}{H_-},\qquad
A=Vdt+\psi(d\varphi-\omega dt),\notag\\
\omega&=\frac{2ma\rho }{\mathscr A}+\frac{{E_0^{2}B_0^{2}}K_K}{4},\notag\\
V&=\frac{E_0}{2}\left(h_K+\frac{{E_0B_0} K_K}{2}\right),\notag\\
\psi&=\frac{-2E_0\mathcal{W}(1-{E_0B_0} \mathcal{W})+B_0s+E_0^2B_0s^2/2}{2H_-}.
\label{eq:fields}\end{align}
The auxiliary functions entering the frame-dragging function and electric potential are
\begin{align}
h_K&=-\frac{2\cos\theta\,\Delta}{\mathscr A}
\bigl[\rho^3-a^2m+a^2(\rho-m)\cos^2\theta\bigr],\notag\\
K_K&=\frac{2ma}{\rho}\Biggl\{a^2-3\rho^2
+\frac{\cos^2\theta\,\Delta}{\mathscr A}
\Bigl[-a^4+6a^2m\rho\notag\\
&-3a^2\rho^2-6\rho^4+\cos^2\theta\bigl(2a^2m\rho-3a^2\rho^2+\rho^4\bigr)
\Bigr]\Biggr\}.
\label{eq:app-currents}\end{align}
Systematic derivation of this spacetime is given in~\cite{SupplementalMaterial}.

To describe the horizons and symmetry axes, define
\begin{align}
&c=\sqrt{m^2-a^2},\quad \rho_\pm=m\pm c,\quad C=|1-\zeta^2|\notag\\
&\mathcal R_+=\rho_+^2+a^2=2m\rho_+,\quad
\zeta=2maE_0B_0 .
\label{eq:horizon-definitions}
\end{align}
The outer horizon lies at $\rho=\rho_+$, and $\mathcal R_+$ sets its area, temperature and angular-velocity scales. The dimensionless $\zeta$ couples rotation to the external-field product; $C$ fixes axial regularity and the azimuthal period.

\emph{Regular-exterior statement.} For finite parameters satisfying \(m>|a|\), \(E_0B_0\neq0\), and \(C>0\), the solution describes a regular exterior bounded by a compact, nondegenerate Killing horizon at \(\rho=\rho_+\). Throughout the open exterior, \(H_\pm\) are positive. On both symmetry axes, they give \(H=C\), so the conical singularities can be removed by introducing the \(2\pi\)-periodic angle \(\widehat{\varphi}\), with \(\varphi=C\widehat{\varphi}\). The horizon admits a regular ingoing chart (see Ref.~\cite{RaczWald1992} for local extensions of Killing horizons), and the electromagnetic potential is regular in separate north and south gauge patches. Consequently, regular exterior solutions exist in all three ranges, \(\zeta<-1\), \(-1<\zeta<1\), and \(\zeta>1\). These intervals define mutually separated subsets of the regular parameter space, but do not by themselves provide a complete classification of its connected components. The black-hole exterior refers specifically to \(\rho>\rho_+\), bounded by the regular future Killing horizon. 

%=============================================================================================================================================================
\section{Thermodynamic degeneracy and intrinsic horizon geometry}
\subsection{Thermodynamics and exact charge balance}
We use the time coordinate $t$ and the $2\pi$-periodic azimuthal angle $\widehat\varphi$. The Killing-horizon generator is $\partial_t+\Omega_H\partial_{\widehat\varphi}$~\cite{BardeenCarterHawking1973}, with
\begin{align}
A_H&=4\pi C\mathcal R_+,\qquad
T_H=\frac{c}{2\pi\mathcal R_+},\notag\\
\Omega_H&=\frac{a}{C\mathcal R_+}
[1-m^2E_0^2B_0^2(3\rho_+^2-a^2)].
\label{eq:horizon}
\end{align}
These quantities use the specified time normalization and angular frame. For the north--south symmetric electromagnetic lift, the quasi-local angular momentum and electromagnetic fluxes are
\begin{align}
J_{\rm sym}&=ma(1+\zeta^2),\notag\\
Q&=2\epsilon_{\rm ax}maB_0,\qquad
P_{\rm mag}=-2\epsilon_{\rm ax}maE_0,
\label{eq:charges}
\end{align}
where $\epsilon_{\rm ax}=\operatorname{sgn}(1-\zeta^2)$. Here $J_{\rm sym}$ is the quasi-local angular momentum for this specified lift (SM Secs.~\ref{sm:ad:C} ~\ref{sm:ad:D}).

The charges and horizon area obey the exact balance identities
\begin{align}
J_{\rm sym}^{\,2}-Q^2P_{\rm mag}^{\,2}
&=m^2a^2C^2,\notag\\
\left(\frac{A_H}{8\pi}\right)^2
&=m^2a^2C^2+C^2mc\mathcal R_+.
\label{eq:balance}
\end{align}
The positive nonextremality term links the horizon scale to rotation and electromagnetic flux. These are identities of the present solution family and specified lift; general results on horizon existence, topology, and area bounds require their respective stability and matter assumptions~\cite{AnderssonMarsSimon2005,GallowaySchoen2006,JaramilloReirisDain2011,GabachClementJaramillo2012}.

\subsection{Identical horizon data and distinct geometries}

The same horizon area and quasi-local charges correspond
to two distinct regular branches. To establish this
correspondence, define
\begin{equation}
\mathcal H=\frac{S}{2\pi},
\qquad
L=\sqrt{J_{\rm sym}^{\,2}-Q^2P_{\rm mag}^{\,2}},
\label{eq:HL-def}
\end{equation}
where $S$ is the Bekenstein--Hawking entropy~\cite{Wald1993}. We orient the axial generator so that $J_{\rm sym}>0$.
For $J_{\rm sym}>L>0$ and $\mathcal H>L$, at fixed $(S,J_{\rm sym},Q,P_{\rm mag})$, the same horizon
area and conserved charges admit two sets of solution parameters $(m,a,E_0,B_0)$
and axial regularity factors:
\begin{equation}
\begin{aligned}
p_\pm &= \frac{J_{\rm sym}\pm L}{2},
&
C_\pm &= \frac{2L}{J_{\rm sym}\pm L},
\\
m_\pm^2 &= \frac{\mathcal H^2+L^2}{2C_\pm \mathcal H},
&
a_\pm &= \frac{p_\pm}{m_\pm},
\\
E_{0,\pm} &= -\frac{P_{\rm mag}}{2\epsilon_\pm p_\pm},
&
B_{0,\pm} &= \frac{Q}{2\epsilon_\pm p_\pm}
\end{aligned}
\label{eq:two-branch-reconstruction}
\end{equation}
The upper and lower signs correspond to
$\epsilon_\pm=\pm1$ and
$|\zeta_+|<1$, $|\zeta_-|>1$, respectively.

The two branches share more than their area and charges. We use one reference prescription: the common axis clock, the rotational frame
\begin{equation}
\xi_{\rm axis}=C^{-1/2}(\partial_t+\nu_{\rm axis}\partial_{\widehat\varphi}),
\quad \nu_{\rm axis}=-\frac{3m^2aE_0^2B_0^2}{C},
\label{eq:axis-frame}
\end{equation}
and the same electric and magnetic lift rules and additive energy zero. A direct adjusted surface-charge calculation on this finite solution family gives (SM Sec.~\ref{sm:ad:D})
\begin{equation}
E_{\rm axis}=m\sqrt C,\qquad
E_{\rm axis}^2=\frac{S}{4\pi}
+\frac{\pi(J_{\rm sym}^{\,2}-Q^2P_{\rm mag}^{\,2})}{S}.
\label{eq:axis-energy}
\end{equation}
The associated first law coefficients
$(T,\Omega,\Phi_Q,\Phi_P)$ are derivatives of this
energy with respect to $(S,J_{\rm sym},Q,P_{\rm mag})$.
The surface charge calculation independently reproduces
these coefficients from the horizon fields, including
the magnetic work term. Consequently,
\begin{equation}
(T,\Omega,\Phi_Q,\Phi_P)_+
=
(T,\Omega,\Phi_Q,\Phi_P)_- .
\label{eq:equal-readouts}
\end{equation}
Here $T=T_H/\sqrt{C}$ and
$\Omega=(\Omega_H-\nu_{\rm axis})/\sqrt{C}$.
The original coordinate temperatures need not agree.

For the four-parameter family of exact solutions
considered here, the energy is consistently defined
by the corresponding surface charges. This relation
holds for variations of all four solution parameters;
we do not claim that it extends to arbitrary changes
of the boundary data or to a general space of field
configurations. Both branches use the same prescription
for the electromagnetic flux, boundary reference, and
energy zero, although the numerical reference
coefficients may differ. Thus, the two branches can
have identical thermodynamic readouts without having
identical external boundary sources or representing
coexisting phases.

\subsection{Strict splitting of the horizon geometry}
On a spatial horizon section, the intrinsic metric is
\begin{align}
h_H&=\Sigma_HH_H\,d\theta^2
+(C\mathcal R_+)^2f(\theta)d\widehat\varphi^2,\notag\\
f(\theta)&=\frac{\sin^2\theta}{\Sigma_HH_H},
\label{eq:shape}
\end{align}
where $\Sigma_H=\Sigma(\rho_+,\theta)$ and $H_H=H(\rho_+,\theta)$. Both branches have the same area radius $\rho_A=\sqrt{C\mathcal R_+}$ and area form $\rho_A^2\sin\theta\,d\theta\,d\widehat\varphi$. Thus $u=(1+\cos\theta)/2$ identifies the same area fraction, measured from the south pole. For every regular nonextremal pair satisfying the
conditions above, an exact positive factorization gives
\begin{equation}
0<f_{\rm super}(\theta)<f_{\rm sub}(\theta),
\qquad 0<\theta<\pi.
\label{eq:strict-geometry-splitting}
\end{equation}
Equation~\eqref{eq:strict-geometry-splitting} compares the subcritical ($|\zeta|<1$) and supercritical ($|\zeta|>1$) reconstructions at the same $(S,J_{\rm sym},Q,P_{\rm mag})$, with $J_{\rm sym}>L>0$ and $\mathcal H>L$. It holds on every nonpolar latitude of these regular nonextremal horizons, including for balanced nonzero fields. At the poles both profiles vanish. The regular extremal limit gives coincident metrics, and the singular critical case $C=0$ and the single-field limits lie outside the two-branch comparison (SM Sec.~\ref{sm:branch:strict-geometry}).

\begin{figure}[!tbp]
\centering
\includegraphics[width=\columnwidth]{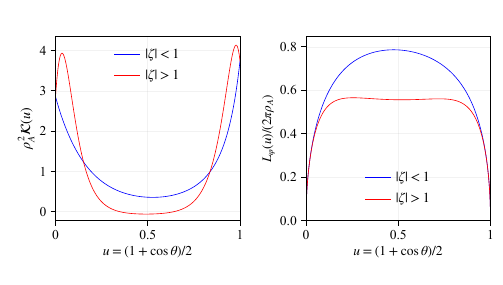}
\caption{Gaussian curvature (left) and latitude-circle circumference (right) for two horizons of equal area. Both panels use the common area fraction $u$, measured from the south pole. The subcritical solution has $(m,a,E_0,B_0)=(1,0.6,0.3,0.5)$ and the red curve is its reconstructed supercritical partner [Eq.~\eqref{eq:two-branch-reconstruction}]. The curvatures agree at corresponding poles. The strict ordering concerns the circumference: at every $0<u<1$ the supercritical circle is shorter.}
\label{fig:branches}
\end{figure}

Since $L_\varphi=2\pi\rho_A^2\sqrt f$, each nonpolar latitude circle is shorter on the supercritical branch. The meridional line element $d\ell_\theta=\sin\theta\,d\theta/\sqrt f$ gives a longer meridian between any two distinct corresponding latitudes. In particular, the two horizons have different maximal orbit circumferences and pole-to-pole distances despite their identical areas and thermodynamic readouts (Fig.~\ref{fig:branches}).

Let $\Lambda^{(0)}_{n,b}$ denote the ordered axisymmetric scalar eigenvalues of $-\Delta_{h_b}$ on either horizon. For the same regular pair, the shape inequality~\eqref{eq:strict-geometry-splitting} orders every positive eigenvalue in this sector:
\begin{equation}
0<\Lambda^{(0)}_{n,{\rm super}}<\Lambda^{(0)}_{n,{\rm sub}},
\qquad n\ge1 .
\label{eq:axisymmetric-order}
\end{equation}
Thus, although the two black holes share identical
specified thermodynamic readouts, all their positive
axisymmetric scalar-Laplacian eigenvalues are strictly
different, showing that the intrinsic horizon spectrum
distinguishes geometries that thermodynamic observables
alone cannot distinguish.~\ref{sm:ad:axisymmetric-spectrum}~\cite{AbreuFreitas2002}.

\section{Global asymptotic matching}
\label{sec:global-matching}

\subsection{Two rotating cylindrical backgrounds and their overlap}
In Weyl coordinates $\varpi=\sqrt{\Delta}\sin\theta$ and $z=(\rho-m)\cos\theta$, the northern ($\sigma=+1$) and southern ($\sigma=-1$) polar regions of the exterior end select the reference metrics
\begin{equation}
\begin{aligned}
ds_{b,\sigma}^2={}&H_{b,\sigma}(-dt^2+d\varpi^2+dz^2)\\
 &+\frac{\varpi^2}{H_{b,\sigma}}(d\varphi-\omega_{b,\sigma}dt)^2,\\
\omega_{b,\sigma}={}&-4\sigma maE_0^2B_0^2 z+m^2aE_0^2B_0^2 .
\end{aligned}
\label{eq:cylindrical-metric}
\end{equation}
Here $H_{b,\sigma}$ is the function $H$ in Eq.~\eqref{eq:field-factors} evaluated at $s=\varpi^2$ and $\mathcal{W}=2\sigma ma$. Both references have angular period $2\pi C$. For $aE_0B_0\ne0$ and $C>0$, their axial dilaton values differ and their dragging functions have opposite slopes in $z$. A single rotating cylindrical background cannot provide the asymptotic description of both polar regions. SM Sec.~\ref{sm:eh:m:sec:extract} extracts the two limits from the black-hole solution, and independently verifies the full four-dimensional EMD equations for each background~\ref{sm:eh:m:sec:verify}.

\begin{figure}[htbp]
\centering
\includegraphics[width=\columnwidth]{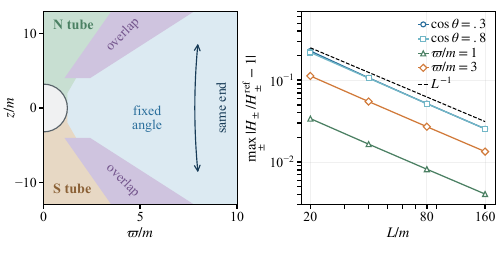}
\caption{Two exact rotating cylindrical references and
their asymptotic compatibility. Left: The northern and
southern references describe overlapping angular regions
of one connected spatial end. Right: Relative differences
between the full-solution factors $H_\pm$ and their
corresponding reference values along fixed-angle and
polar-tube paths, showing their asymptotic convergence.
The full solution is smooth across the overlap, while
the two reference backgrounds become asymptotically
compatible there.}
\label{fig:matchingatlas}
\end{figure}

On every fixed angular overlap $|\cos\theta|\le1-\delta$, $0<\delta<1$, the references are compatible. The relative metric factor, dilaton difference, and rotational shear between their orthonormal frames satisfy
\begin{equation}
\begin{aligned}
H_{b,+}/H_{b,-}-1&=O(\rho^{-4}),\\
\phi_{b,+}-\phi_{b,-}&=O(\rho^{-4}),\\
\varpi(\omega_{b,+}-\omega_{b,-})/H_{b,+}&=O(\rho^{-2}).
\end{aligned}
\label{eq:background-overlap}
\end{equation}
For fixed $m>|a|$, $E_0B_0\ne0$ and $C>0$, there are constants $K$ and $\rho_0$ such that on either closed hemisphere, for $\rho\ge\rho_0$,
\begin{equation}
\sup_{\sigma\cos\theta\ge0}\mathcal N_\sigma\le K\frac m\rho,
\qquad
\sup_{\sigma\cos\theta\ge0}|\beta_\sigma|\le K\frac{m^2}{\rho^2}.
\label{eq:uniform-matching}
\end{equation}
Here $\mathcal N_\sigma$ measures the metric, dilaton and Maxwell differences and their weighted first orbit-space derivatives in the reference frame. The rotational mismatch is $\beta_\sigma=\varpi(\omega-\omega_{b,\sigma})/H_{b,\sigma}$. SM Sec.~\ref{sm:eh:m:sec:matching} defines the norm and proves the overlap and uniform estimates, including the axes. The two references describe one connected end of the exact exterior (Fig.~\ref{fig:matchingatlas}).

%=============================================================================================================================================================

\subsection{Global hyperbolicity of the regular exterior}
Every regular noncritical exterior $\mathcal M_{\rm ext}=\{\rho>\rho_+\}$ is globally hyperbolic, including the supercritical branches. Direct inversion gives
\begin{equation}
g^{tt}=-\frac{\mathscr A}{H\Sigma\Delta}<0,
\label{eq:temporal-condition}
\end{equation}
so $t$ is a time function. Every future-directed causal tangent satisfies
\begin{equation}
dt\geq\frac{\sqrt{\mathscr A}}{\Delta}|d\rho|
\geq\frac{\rho^2}{\Delta}|d\rho|,
\qquad \rho>\rho_+.
\label{eq:cone}
\end{equation}
The integral of $\rho^2/\Delta$ diverges at both radial ends. If $t$ were bounded above on a future-inextendible causal curve, Eq.~\eqref{eq:cone} would confine its tail to a closed radial interval inside the exterior. Together with the bounded time interval and the regular angular sphere, this gives a compact spacetime set. Since $t$ is temporal, strong causality excludes total imprisonment there. SM Sec.~\ref{sm:eh:c:sec:causal} gives the compactness argument in both time directions. Thus each constant-$t$ slice is a Cauchy surface of topology $(\rho_+,\infty)\times S^2$~\cite{Geroch1970,BernalSanchez2005,CaponioJavaloyesSanchez2011}.

\subsection{Global causal structure}

Every regular noncritical exterior
$\mathcal M_{\rm ext}=\{\rho>\rho_+\}$ is globally
hyperbolic, including the supercritical branches.
Indeed, direct inversion gives
\begin{equation}
g^{tt}=-\frac{\mathscr A}{H\Sigma\Delta}<0,
\qquad
dt\geq\frac{\rho^2}{\Delta}|d\rho|.
\label{eq:global-hyperbolicity}
\end{equation}
The associated radial optical distance diverges at
both $\rho=\rho_+$ and $\rho\to\infty$. Together with
the compact angular sections, this establishes that
the constant-$t$ hypersurfaces are Cauchy surfaces.

We further identify the future outer event horizon
relative to the specified axial future escape end.
Let $\gamma_u$ denote the family of outgoing axial
null geodesics defined by $t-\rho_* =u$, with
$d\rho_*/d\rho=\mathcal R/\Delta$. In the selected
future-ingoing extension, their causal past satisfies
\begin{equation}
\bigcup_{u\in\mathbb R}
I^-_{\mathcal M_{\rm in}}(\gamma_u)
=
\{\rho>\rho_+\},
\quad
\mathcal H^+_{\rm out}
=
\{\rho=\rho_+\}_{\rm future}.
\label{eq:global-event-horizon}
\end{equation}
Thus the outer Killing horizon is also the global
future event horizon relative to this specified
escape end. 

\section{Conclusion}

We have constructed an exact finite-rotation
KK black-hole solution of four-dimensional
EMD theory,
immersed in independent external electric and magnetic
fields. Its horizon
observables admit two distinct regular nonextremal
parameter reconstructions, governed by the exact
balance identities and inverse relations
(\hyperref[eq:balance]{3.3},
\hyperref[eq:two-branch-reconstruction]{3.5}).
Under a common, explicitly specified reference
prescription for the axial clock, rotational frame,
and electromagnetic potentials, the two branches
have identical thermodynamic readouts
\[
(S,J_{\rm sym},Q,P_{\rm mag},
E_{\rm axis},T,\Omega,\Phi_Q,\Phi_P),
\]
as established by
(\hyperref[eq:axis-energy]{3.7},
\hyperref[eq:equal-readouts]{3.8}), despite having
different solution parameters. The energy follows
from an integrable surface-charge variation along
the four independent parameters of the exact solution
family, with the magnetic-flux polarization and
boundary reference fixed as specified. This equality
holds within the considered solution family, while
the two branches can have different external-field
boundary sources.

The geometric distinction follows from the exact
inequality
(\hyperref[eq:strict-geometry-splitting]{3.10}),
which gives shorter latitude circles and longer
corresponding meridional segments on the supercritical
branch. Their positive axisymmetric scalar-Laplacian
eigenvalues are also strictly ordered
(\hyperref[eq:axisymmetric-order]{3.11}). Thus, the
two black holes have identical specified thermodynamic
readouts but different intrinsic horizon geometries
and scalar spectra. The geometric spectrum therefore
distinguishes horizons that cannot be distinguished
by the specified thermodynamic observables alone.
The two terms in the axisymmetric spectral operator
contribute oppositely to the comparison, so we cannot
infer from this result how the non-axisymmetric spectrum
is ordered or whether either branch is dynamically
stable.

We have also shown that the exterior admits two
distinct exact rotating cylindrical backgrounds
associated with its northern and southern polar regions.
Their metric, dilaton, and electromagnetic fields become
asymptotically compatible in the angular overlap region,
with uniform matching extending to both symmetry axes
(\hyperref[eq:background-overlap]{4.2},
\hyperref[eq:uniform-matching]{4.3}). Every regular
noncritical exterior, including the supercritical
branches, is globally hyperbolic and admits Cauchy
surfaces of constant $t$. We further identify the
future outer Killing horizon as the global event
horizon relative to the specified axial future escape
end (\hyperref[eq:global-event-horizon]{4.7}).

\begin{acknowledgments}
This work is supported by the National Natural Science Foundation of China
(NSFC) under Grant Nos.~12373022 and U1731107.
\end{acknowledgments}

\onecolumngrid

%=============================================================================================================================================================
% Begin merged source: sm_reference_keys.tex

% End merged source: sm_reference_keys.tex

% Verified reference list; synchronized across both languages and the Supplemental Material.
%

% BEGIN APPENDED SUPPLEMENTAL MATERIAL
\clearpage
\onecolumngrid
\setlength{\emergencystretch}{3em}
\setlength{\jot}{2pt}
\allowdisplaybreaks[2]\raggedbottom
\setcounter{secnumdepth}{2}
\counterwithout*{equation}{section}
\setcounter{section}{0}\setcounter{subsection}{0}\setcounter{subsubsection}{0}
\setcounter{equation}{0}\setcounter{figure}{0}\setcounter{table}{0}
\renewcommand{\thesection}{\Alph{section}}
\renewcommand{\thesubsection}{\alph{subsection}}
\renewcommand{\theequation}{S\arabic{equation}}
\renewcommand{\thefigure}{S\arabic{figure}}
\renewcommand{\thetable}{S\arabic{table}}
\renewcommand{\theHsection}{SM.\arabic{section}}
\renewcommand{\theHequation}{SM.\arabic{equation}}
\renewcommand{\theHfigure}{SM.\arabic{figure}}
\renewcommand{\theHtable}{SM.\arabic{table}}
\hypertarget{supplemental-material}{}
\begin{center}
\bfseries
Supplemental Online Material for\\[0.5ex]
``Thermodynamic Degeneracy and Distinct Horizon Geometries of Rotating Kaluza–Klein\\ Black Holes in Independent External Electric and Magnetic Fields''
\end{center}
\begin{quotation}
\noindent In this Supplemental Material, we verify the
exact solution against the four-dimensional EMD field
equations, derive the two-branch thermodynamic
reconstruction and Hamiltonian integrability, establish
the strict horizon-geometry inequality and axisymmetric
spectral splitting, and prove the exactness and uniform
asymptotic matching of the two cylindrical backgrounds,
together with the global hyperbolicity of the open
exterior and the identification of its future outer
event horizon relative to the specified axial escape end.
\end{quotation}
\twocolumngrid

\section{Field-equation verification and independent checks}

\label{sm:ad:B}

The four-dimensional field equations following from the action~\mainEq{eq:action}{2.1} are

\begin{align}
&R_{\mu\nu}=2\nabla_\mu\phi\nabla_\nu\phi
 +2e^{-2\sqrt3\phi}\left(F_{\mu\lambda}F_\nu{}^\lambda
 -\frac14g_{\mu\nu}F^2\right),\notag\\
&\nabla_\mu\!\left(e^{-2\sqrt3\phi}F^{\mu\nu}\right)=0,\quad \nabla^2\phi+\frac{\sqrt3}{2}e^{-2\sqrt3\phi}F^2=0.
\label{sm:ad:emd-equations}
\end{align}

We regard the metric, Maxwell potential and dilaton in Eqs.~\mainEq{eq:metric}{2.2}--\mainEq{eq:app-currents}{2.6} as given four-dimensional fields. Their substitution into Eq.~\eqref{sm:ad:emd-equations} uses
\begin{align}
&\sqrt{-g}=H\Sigma\sin\theta,\quad g^{\rho\rho}=\frac{\Delta}{H\Sigma},\quad g^{\theta\theta}=\frac{1}{H\Sigma},\notag\\
&g^{tt}=-\frac{\mathscr A}{H\Sigma\Delta},\quad g^{t\varphi}=\omega g^{tt},\quad g^{\varphi\varphi}=\frac{H}{s}+\omega^2g^{tt}.
\label{sm:ad:direct-inverse}
\end{align}
For $i=\rho,\theta$, the nonzero field-strength components and Maxwell flux densities are
\begin{align}
&F_{it}=\partial_iV-\psi\partial_i\omega-\omega\partial_i\psi,\quad F_{i\varphi}=\partial_i\psi,\notag\\
&\mathcal J_i^t=\sqrt{-g}\,e^{-2\sqrt3\phi}g^{ii}g^{tt}(\partial_iV-\psi\partial_i\omega),\notag\\
&\mathcal J_i^\varphi=\omega\mathcal J_i^t+
\sqrt{-g}\,e^{-2\sqrt3\phi}g^{ii}\frac{H}{s}\partial_i\psi,
\label{sm:ad:direct-maxwell-inputs}
\end{align}
with no sum over $i$. Thus the two nontrivial Maxwell equations are
$\partial_\rho\mathcal J_\rho^a+\partial_\theta\mathcal J_\theta^a=0$, $a=t,\varphi$;
the other two components vanish by stationarity and axial symmetry. The Bianchi identity follows from $F=dA$. The scalar Laplacian is
\begin{equation}
\nabla^2\phi=\frac{1}{H\Sigma}
\left[\partial_\rho(\Delta\partial_\rho\phi)
+\frac{1}{\sin\theta}\partial_\theta(\sin\theta\partial_\theta\phi)\right].
\label{sm:ad:direct-scalar}
\end{equation}
We calculate $R_{\mu\nu}$ from the four-dimensional Levi--Civita connection of the displayed metric. Write $\mathcal E_{\mu\nu}$ for the difference between the two sides of the Einstein equation in Eq.~\eqref{sm:ad:emd-equations}, $\mathcal M^a=\partial_i\mathcal J_i^a$, and $\mathcal E_\phi=\nabla^2\phi+(\sqrt3/2)e^{-2\sqrt3\phi}F^2$. Substitution of the explicit functions and cancellation gives
\begin{align}
&\mathcal E_{tt}=\mathcal E_{t\varphi}=\mathcal E_{\varphi\varphi}
=\mathcal E_{\rho\rho}=\mathcal E_{\rho\theta}=\mathcal E_{\theta\theta}=0,\notag\\
&\mathcal M^t=\mathcal M^\varphi=0,\quad \mathcal E_\phi=0.
\label{sm:ad:direct-emd-identities}
\end{align}
The mixed Einstein components between $(t,\varphi)$ and $(\rho,\theta)$ vanish identically. For the algebraic cancellation, $H^2=H_+H_-$ and the logarithmic derivatives of $H$ and $\phi$ eliminate square roots from the independent residuals after multiplication by nonzero metric factors. After clearing the nonvanishing denominators, the independent residual numerators vanish identically as polynomials in $\rho$ and $\cos\theta$, with coefficients depending on $m,a,E_0,B_0$. These identities establish the full EMD equations on the chosen real branch $H_\pm>0$ wherever the given fields and coordinates are regular.

\section{Horizon, common axial period, and gauge patches}
\label{sm:ad:C}

On an axis \(\cos\theta =\sigma=\pm1\),
\begin{align}
&s=0,\quad \mathcal{W}=2\sigma ma,\quad H_{+,\sigma}=(1+\sigma\zeta)^2,\notag\\
&H_{-,\sigma}=(1-\sigma\zeta)^2,\quad H_\sigma=C.
\label{sm:ad:axis-values}
\end{align}
Both exterior axes and both horizon poles are smoothed by the same positive
period
\begin{equation}
\varphi=C\widehat\varphi,\quad \widehat\varphi\sim\widehat\varphi+2\pi.
\label{sm:ad:common-period}
\end{equation}
The two critical surfaces \(\zeta=\pm1\) are excluded.  The regular ranges on
both sides use \(C\), not the signed factor \(1-\zeta^2\).

At the outer horizon,
\begin{equation}
h_K\big|_{\rho _+}=0,\quad K_K\big|_{\rho _+}=\frac{2ma}{\rho _+}(a^2-3\rho _+^2),
\label{sm:ad:horizon-current}
\end{equation}
so \(\omega_H\) and \((A_t+\omega_H\psi)_H\) are independent of \(\theta\).
Using the $2\pi$-periodic angle of Eq.~\eqref{sm:ad:common-period} gives the observables in Eq.~\mainEq{eq:horizon}{3.1} and

\begin{equation}
\kappa_H=\frac{c}{\mathcal R_+},\quad 2T_H\frac{A_H}{4}=Cc.
\label{sm:ad:horizon-observables}
\end{equation}

A future-horizon chart is obtained from
\begin{equation}
dv=dt+\frac{\mathcal R_+}{\Delta}\,d\rho ,\quad d\widetilde\varphi=d\varphi
 +\frac{\omega_H\mathcal R_+}{\Delta}\,d\rho .
\label{sm:ad:ingoing-chart}
\end{equation}
Since \(\mathscr A-\mathcal R_+^2=O(\rho -\rho _+)\) and
\(\omega-\omega_H=O(\rho -\rho _+)\), the metric is regular in this chart.  A
constant shift of the horizon electric potential removes the radial Maxwell
gauge pole.

Magnetic flux requires two axial gauge patches.  With
\(\psi_\sigma=\psi|_{\cos\theta =\sigma}\), take
\begin{align}
&A_N=A_tdt+C(\psi-\psi_N)d\widehat\varphi,\notag\\
&A_S=A_tdt+C(\psi-\psi_S)d\widehat\varphi.
\label{sm:ad:gauge-patches}
\end{align}
On their overlap,
\begin{equation}
A_S-A_N
 =C(\psi_N-\psi_S)d\widehat\varphi
 =2P_{\rm mag}\,d\widehat\varphi.
\label{sm:ad:gauge-transition}
\end{equation}
The sign follows from the orientation convention
\(-dt\wedge d\rho \wedge d\theta\wedge d\varphi>0\).  The corresponding electric and
magnetic fluxes are
\begin{equation}
Q=\frac{C}{2}(\chi_N-\chi_S),\quad P_{\rm mag}=\frac{C}{2}(\psi_N-\psi_S).
\label{sm:ad:flux-endpoints}
\end{equation}
The transition in Eq.~\eqref{sm:ad:gauge-transition} must be accompanied by
the same lift of the axial rotation in the Maxwell bundle when angular
momentum is evaluated.

\section{Quasi-local angular momentum and charge balance}
\label{sm:ad:D}

We use the north--south symmetric Maxwell lift to
define the quasi-local angular momentum
$J_{\rm sym}$~\cite{Prabhu2017}. The corresponding
stationary-slice Noether integral depends on the
choice of lift and is not an ADM angular momentum
\cite{IyerWald1994,GibbonsPangPope2014}.

For the $2\pi$-periodic axial generator
$\eta=\partial_{\widehat\varphi}$, let
$G=e^{-2\sqrt3\phi}{}^\star F$. In each regular
electromagnetic gauge patch, the Maxwell lift is
represented by a constant gauge parameter $\lambda$
and the combination
\begin{equation}
\mu=\iota_\eta A+\lambda,
\qquad
d\mu=-\iota_\eta F.
\label{sm:ad:lift-definition}
\end{equation}
Fixing this lift and pulling the Noether two-form back
to a stationary spatial slice, its on-shell integral
over homologous two-surfaces in the same regular gauge
patch is~\cite{IyerWald1994,GibbonsPangPope2014}
\begin{equation}
J_\mu[S]
=
\frac{1}{16\pi}\int_S
\left({}^\star d\eta^\flat+4\mu G\right).
\label{sm:ad:surface-charge}
\end{equation}

The north--south symmetric choice is
\begin{align}
\mu_{\rm sym}
&=
C\left[
\psi-\frac{\psi_N+\psi_S}{2}
\right],
\notag\\
\lambda_N&=P_{\rm mag},
\qquad
\lambda_S=-P_{\rm mag}.
\label{eq:symmetric-lift-SM}
\end{align}
Using the gauge patches in
Eq.~\eqref{sm:ad:gauge-patches} and
$P_{\rm mag}=C(\psi_N-\psi_S)/2$, both
$\iota_\eta A_N+\lambda_N$ and
$\iota_\eta A_S+\lambda_S$ equal
$\mu_{\rm sym}$. Thus the same lift is used on
the overlap. Substitution of the exact fields into
Eq.~\eqref{sm:ad:surface-charge}, together with
the endpoint fluxes in Eq.~\eqref{sm:ad:flux-endpoints},
gives the charges in
Eq.~\mainEq{eq:charges}{3.2}.

For a spacetime-constant shift $k$ of the lift,
the surface-charge formula and
$Q=(4\pi)^{-1}\int_S G$ give
\begin{equation}
J_{\mu+k}=J_\mu+kQ.
\label{sm:ad:lift-shift}
\end{equation}
The north-patch, south-patch, and unshifted choices are
\[
\mu_N=C(\psi-\psi_N),\qquad
\mu_S=C(\psi-\psi_S),\qquad
\mu_{\rm raw}=C\psi.
\]
Their stationary-slice integrals are
\begin{align}
J_N&=ma(1+\zeta)^2,\qquad
J_S=ma(1-\zeta)^2,\notag\\
J_{\rm raw}&=ma(1-\zeta^2).
\label{sm:ad:other-lifts}
\end{align}
In particular,
\[
\begin{aligned}
J_N-J_{\rm sym}&=-QP_{\rm mag}=2ma\zeta,\\
J_S-J_{\rm sym}&=QP_{\rm mag}=-2ma\zeta.
\end{aligned}
\]
These differences show that the angular momentum
entering the balance identities is tied to the
specified symmetric lift. A change of lift shifts
the quasi-local angular momentum and cannot be
ignored when comparing the two branches.

To verify the exact balance identities, we use
\[
QP_{\rm mag}=-2ma\zeta,
\qquad
(1+\zeta^2)^2-4\zeta^2=C^2.
\]
The angular-momentum identity in
Eq.~\mainEq{eq:balance}{3.3} then follows directly.
The area identity follows from
Eq.~\mainEq{eq:horizon}{3.1} and
\[
\frac{\mathcal R_+^2}{4}-m^2a^2
=mc\mathcal R_+.
\]
Thus both balance relations are verified using
the same quasi-local angular-momentum prescription
that enters the two-branch reconstruction.

\subsection{Identical horizon data and distinct geometries}

We provide the algebraic derivation and consistency
conditions underlying the two-branch reconstruction
given in the main text. We use the horizon variables
$\mathcal H$ and $L$ defined there, and set
$p=ma>0$.

The charge relations imply
\begin{align}
&L^2
=J_{\rm sym}^{\,2}-Q^2P_{\rm mag}^{\,2}\quad =p^2\left[(1+\zeta^2)^2-4\zeta^2\right]\notag\\
&=p^2(1-\zeta^2)^2.
\label{eq:branch-L-identity-SM}
\end{align}
Since $C=|1-\zeta^2|$, this gives
\begin{equation}
L=pC.
\label{eq:branch-LpC-SM}
\end{equation}
The two signs in the main-text reconstruction therefore
correspond to the two regular parameter ranges
$|\zeta|<1$ and $|\zeta|>1$, respectively.

To verify the reconstruction of the external fields,
the inverse charge relations give
\begin{equation}
E_{0,\pm}
=-\frac{P_{\rm mag}}{2\epsilon_\pm p_\pm},\quad B_{0,\pm}
=\frac{Q}{2\epsilon_\pm p_\pm},\quad \epsilon_\pm=\pm1.
\label{eq:branch-fields-SM}
\end{equation}
Consequently,
\begin{equation}
2p_\pm E_{0,\pm}B_{0,\pm}
=-\frac{QP_{\rm mag}}{2p_\pm}
=\zeta_\pm,\quad C_\pm=|1-\zeta_\pm^2|.
\label{eq:branch-field-consistency-SM}
\end{equation}
Thus, the reconstructed amplitudes reproduce the
original electric and magnetic fluxes on each branch.

The solution parameters are checked using the exact
horizon identities. In particular,
\begin{align}
&\mathcal H^2+L^2
=C^2m^2(\rho_+^2+a^2)
=2C\mathcal H m^2,\notag\\
&\mathcal H^2-L^2
=C^2m^2(\rho_+^2-a^2)
=2C\mathcal H m c,
\label{eq:branch-seed-identities-SM}
\end{align}
where we used
\[
\mathcal H=Cm\rho_+,\qquad
\rho_+^2-a^2=2c\rho_+.
\]
These identities establish the reality and
subextremality of the reconstructed solution parameters.
In particular, for
\[
\mathcal H>L>0,\qquad J_{\rm sym}>L,
\]
both branches have
\[
m_\pm>|a_\pm|,\qquad c_\pm>0.
\]
The condition $J_{\rm sym}>L$ is satisfied for
nonzero electric and magnetic fluxes with
$J_{\rm sym}>0$.

For the original coordinate clock, $c_H=2T_HS=Cc$. The horizon identities give
\begin{equation}
c_{H,\pm}^{\,2}
=
\frac{L(\mathcal H^2-L^2)^2}
{\mathcal H(\mathcal H^2+L^2)(J_{\rm sym}\pm L)}.
\label{eq:horizon-two-branch-closure}
\end{equation}

\subsection{Axis-clock prescription and equal thermodynamic readouts}

We specify the boundary prescription before identifying an energy. On the finite-dimensional exact family with $m>|a|>0$, $C>0$ and $QP_{\rm mag}\ne0$, we use a classical real magnetic-flux polarization and a relative edge reference. The resulting energy is an adjusted surface Hamiltonian on this finite-dimensional family, not an ADM mass. Covariant surface charges and electromagnetic lifts provide the framework~\cite{LeeWald1990,IyerWald1994,GibbonsPangPope2014,Prabhu2017}; the prescription and formulas below apply to the present solution family.

For completeness, its magnetic-flux realization can be given on an exterior annulus. Let $\boldsymbol\beta=d(\cos\theta)\wedge d\widehat\varphi$, with $\int_{S^2}\boldsymbol\beta=4\pi$ in the flux orientation used above. The reference form $\boldsymbol\beta$ is fixed under field variations, and the time and axial generators used here preserve it, $\mathcal L_\xi\boldsymbol\beta=0$. Introduce a global regular one-form $\mathcal A_{\rm reg}$ and an auxiliary three-form $K_3$:
\begin{align}
F&=d\mathcal A_{\rm reg}+P_{\rm mag}\boldsymbol\beta,\notag\\
I_{\rm flux}&=I_{\rm EMD}[g,\phi,F]
 +\frac1{4\pi}\int K_3\wedge dP_{\rm mag},\notag\\
\Theta_{\rm flux}&=\Theta_{g,\phi}
 -\frac1{4\pi}(\delta\mathcal A_{\rm reg}\wedge G+K_3\delta P_{\rm mag}).
\label{sm:ad:axis-flux-polarization}
\end{align}
Variation gives $dP_{\rm mag}=0$, $dG=0$ and $dK_3=\boldsymbol\beta\wedge G$, preserving the original local EMD equations. The equivalent topological polarization has edge symplectic form $\delta P_{\rm mag}\wedge\delta u_m$, where $u_m=(4\pi)^{-1}\int_\Sigma K_3$. A relative transformation $K_3\mapsto K_3+d\Lambda$, with $\Lambda=-\lambda_m\boldsymbol\beta$ at the outer sphere and zero at the inner sphere, gives $\delta_\Lambda u_m=-\lambda_m$ and surface variation $-\lambda_m\delta P_{\rm mag}$. This is a real-flux phase space, rather than continuous variation within one fixed compact $U(1)$ bundle.

Write $\eta=\partial_{\widehat\varphi}$ and $db_\xi=-\iota_\xi\boldsymbol\beta$, with $b_{\partial_t}=0$ and $b_\eta=\cos\theta$. The Noether two-form is
\begin{equation}
\boldsymbol Q_{\xi,\lambda_e}
=-\frac{{}^\star d\xi^\flat}{16\pi}
-\frac{\iota_\xi\mathcal A_{\rm reg}+P_{\rm mag}b_\xi+\lambda_e}{4\pi}G.
\label{sm:ad:axis-flux-noether}
\end{equation}
Taking $\mathcal A_{\rm reg}=A_tdt+[C(\psi-\bar\psi)-P_{\rm mag}\cos\theta]d\widehat\varphi$, where $\bar\psi=(\psi_N+\psi_S)/2$, reproduces the specified symmetric axial lift. Its adjusted variation gives $\delta H_{-\eta}=\delta J_{\rm sym}$; an electric constant lift contributes $-\lambda_e\delta Q$. State-dependent generators are varied with the standard generator-variation subtraction, so no terms $-Q\delta\lambda_e$ or $-P_{\rm mag}\delta\lambda_m$ are included.

The electromagnetic horizon coefficients follow from the fields, independently of an energy ansatz. The local dual potential has
\begin{align}
A^d&=V_d dt+\chi(d\varphi-\omega dt),\quad dA^d=G,\notag\\
V_d&=-\frac{B_0}{2}h_K+\frac{E_0B_0^2}{4}K_K,\notag\\
\bar\psi&=-\frac{4m^2a^2E_0^2B_0}{1-\zeta^2},\quad
\bar\chi=-\frac{4m^2a^2E_0B_0^2}{1-\zeta^2}.
\label{sm:ad:axis-dual-and-endpoints}
\end{align}
Let $\omega_H=C\Omega_H$, $V_H=V|_H$ and $V_{d,H}=V_d|_H$. The horizon surface form is obtained by taking the
limit from the regular exterior. This procedure does
not require the auxiliary three-form $K_3$ to remain
regular at the bifurcation horizon. This limit yields
\begin{align}
\Phi_E&=-V_H+\omega_H\bar\psi,\quad
\Phi_P=V_{d,H}-\omega_H\bar\chi,\notag\\
\boldsymbol\vartheta_t&=T_H\delta S+\Omega_H\delta J_{\rm sym}
 +\Phi_E\delta Q+\Phi_P\delta P_{\rm mag}.
\label{sm:ad:axis-source-surface-form}
\end{align}
In particular, the magnetic coefficient is the exact endpoint integral
\begin{equation}
\frac12\int_0^\pi
 [V_{d,H}\sin\theta+\omega_H\partial_\theta(\cos\theta\,\chi_H)]d\theta
=V_{d,H}-\omega_H\bar\chi.
\label{sm:ad:axis-magnetic-endpoint}
\end{equation}
This follows by taking $G=d\mathcal B_{\rm reg}+Q\boldsymbol\beta$
and $K_3=\boldsymbol\beta\wedge\mathcal B_{\rm reg}$
in the surface form. Thus, the magnetic work term is
obtained directly from the surface-charge calculation,
rather than introduced through an assumed first law.

The axis-clock prescription fixes
\begin{align}
\mu_{\rm clk}&=C^{-1/2},\quad
\xi_{\rm axis}=\mu_{\rm clk}(\partial_t+\nu_{\rm axis}\eta),\notag\\
\omega_0&=m^2aE_0^2B_0^2,\quad
\nu_{\rm axis}=-\frac{3\omega_0}{C},\notag\\
g_{\rm clk}&=-\frac{3m\zeta^2}{2C^{3/2}},\quad
\lambda_e=-\frac{g_{\rm clk}}Q,\quad
\lambda_m=-\frac{g_{\rm clk}}{P_{\rm mag}}.
\label{sm:ad:axis-clock-prescription}
\end{align}
The time factor gives unit asymptotic time on the axes, where $g_{tt}\to-C$; The time normalization sets the asymptotic clock on
the axes, where $g_{tt}\to-C$, but does not by itself
fix the rotational frame or the electromagnetic
potentials. In the original north and south patches the total electric lift is $\lambda_i^{\rm tot}=\mu_{\rm clk}\nu_{\rm axis}\lambda_i^{(\eta)}+\lambda_e$, with $\lambda_N^{(\eta)}=P_{\rm mag}$ and $\lambda_S^{(\eta)}=-P_{\rm mag}$.

Using the actual horizon values $h_K|_H=0$ and $K_K|_H=2ma(a^2-3\rho_+^2)/\rho_+$ gives, on both regular branches,
\begin{align}
E_{\rm axis}&=m\sqrt C,\quad
E_{\rm axis}^2=\frac{S}{4\pi}
 +\frac{\pi(J_{\rm sym}^2-Q^2P_{\rm mag}^2)}S,\notag\\
T_{\rm axis}&=\mu_{\rm clk}T_H
 =\frac{S^2-4\pi^2(J_{\rm sym}^2-Q^2P_{\rm mag}^2)}{8\pi E_{\rm axis}S^2},\notag\\
\Omega_{\rm axis}&=\mu_{\rm clk}(\Omega_H-\nu_{\rm axis})
 =\frac{\pi J_{\rm sym}}{E_{\rm axis}S},\notag\\
\Phi_{Q,\rm axis}&=\mu_{\rm clk}\Phi_E-\lambda_e
 =-\frac{\pi QP_{\rm mag}^2}{E_{\rm axis}S},\notag\\
\Phi_{P,\rm axis}&=\mu_{\rm clk}\Phi_P-\lambda_m
 =-\frac{\pi P_{\rm mag}Q^2}{E_{\rm axis}S}.
\label{sm:ad:axis-clock-readouts}
\end{align}
Substitution into the independently obtained surface form gives
\begin{align}
\delta H^{\rm flux}_{\xi_{\rm axis},\lambda_e,\lambda_m}
&=\mu_{\rm clk}(\boldsymbol\vartheta_t-\nu_{\rm axis}\delta J_{\rm sym})
 -\lambda_e\delta Q-\lambda_m\delta P_{\rm mag}\notag\\
&=\delta(m\sqrt C).
\label{sm:ad:axis-clock-integrability}
\end{align}
The equality holds in all four independent
solution-parameter directions, not merely after
differentiation of the observable energy formula.
We choose the same additive energy zero on both
regular branches. This establishes integrability
of the adjusted Hamiltonian on the specified
four-parameter family in the chosen flux-polarization
and relative-edge framework. This establishes integrability within the specified
four-parameter family and boundary prescription,
but does not mean that the same energy definition
applies to other boundary conditions or that
integrability holds for more general field variations.

Equal $(S,J_{\rm sym},Q,P_{\rm mag})$ therefore gives equal axis-clock energy and all four adjusted readouts. The coefficients of the prescription need not be numerically equal on the two branches. With $k=|\zeta_+|$, $C_-=C_+/k^2$ and $\mu_{{\rm clk},-}=k\mu_{{\rm clk},+}$. In contrast, the original coordinate temperatures obey
\begin{equation}
\frac{T_{H,-}}{T_{H,+}}
=\sqrt{\frac{J_{\rm sym}+L}{J_{\rm sym}-L}}>1,
\qquad T_{{\rm axis},-}=T_{{\rm axis},+}.
\label{eq:two-branch-temperature}
\end{equation}
Variations of asymptotic scalar and electromagnetic sources can carry additional work terms~\cite{GibbonsKalloshKol1996,Yazadjiev2013}. The two branches have identical adjusted thermodynamic
readouts under the specified prescription, while their
external boundary sources may differ.

\subsection{Strict splitting of the intrinsic horizon geometry}
\label{sm:branch:strict-geometry}

Consider the two branches at fixed $(S,J_{\rm sym},Q,P_{\rm mag})$, with $J_{\rm sym}>L>0$ and $\mathcal H>L$. In this proof, unlabelled solution parameters refer to the subcritical branch, and a tilde denotes its supercritical partner. Put $\lambda=\zeta^2\in(0,1)$. The reconstruction gives
\begin{align}
&\widetilde m=\sqrt\lambda\,m,\quad
\widetilde a=\sqrt\lambda\,a,\quad
\widetilde E_0=-\frac{E_0}{\lambda},\quad
\widetilde B_0=-\frac{B_0}{\lambda},\notag\\
&\widetilde\rho_+=\sqrt\lambda\,\rho_+,\quad
\widetilde{\mathcal R}_+=\lambda\mathcal R_+,\quad
\widetilde C=\frac C\lambda,\quad
\widetilde\zeta=\frac1\zeta.
\label{sm:branch:scaling-map}
\end{align}
Hence both branches have the same area scale $\rho_A^2=C\mathcal R_+=\widetilde C\widetilde{\mathcal R}_+$. At the same $\theta$, their horizon functions obey $\widetilde\Sigma_H=\lambda\Sigma_H$, $\widetilde s_H=\lambda s_H$ and $\widetilde{\mathcal{W}}_H=\lambda \mathcal{W}_H$.

Define $\mathcal X=\Sigma_HH_H>0$, $\widetilde{\mathcal X}=\widetilde\Sigma_H\widetilde H_H>0$ and $\mathcal Z=(E_0B_0)^2(\mathcal{W}_H^2+s_H^2/4)$. The horizon relation $\rho_+^2-a^2=2c\rho_+$ yields the exact identity
\begin{align}
 \mathcal{W}_H^2+\frac{s_H^2}{4}-4m^2a^2
 &=\frac{c^2\mathcal R_+^2\sin^4\theta}{\Sigma_H},
 \notag\\
 \mathcal Z-\lambda
 &=\frac{E_0^2B_0^2c^2\mathcal R_+^2\sin^4\theta}{\Sigma_H}.
 \label{sm:branch:positive-horizon-identity}
\end{align}
Substitution into the two field factors gives a factorization over the entire angular range:
\begin{align}
\widetilde{\mathcal X}^{\,2}-\mathcal X^2
&=\frac{\Sigma_H^2(1-\lambda)(\mathcal Z-\lambda)}{\lambda^2}\notag\\
&\quad\times[\lambda(E_0^2+B_0^2)s_H+(1+\lambda)(\lambda+\mathcal Z)].
\label{sm:branch:positive-metric-gap}
\end{align}
For $c>0$, $E_0B_0\ne0$ and $0<\theta<\pi$, every factor needed for strict positivity is positive. Taking the positive square roots and using the definition of $f$ in the main text therefore proves
\begin{equation}
\begin{aligned}
f_{\rm sub}-f_{\rm super}
&=\frac{\sin^2\theta\,
(\widetilde{\mathcal X}^{\,2}-\mathcal X^2)}
{\mathcal X\widetilde{\mathcal X}
(\mathcal X+\widetilde{\mathcal X})}>0,\\
&\qquad 0<f_{\rm super}<f_{\rm sub}.
\end{aligned}
\label{sm:branch:strict-shape-order}
\end{equation}
At either pole, $f_{\rm super}=f_{\rm sub}=0$ and $\widetilde{\mathcal X}=\mathcal X=\rho_A^2$. The difference in $\mathcal X$ is $O(\sin^4\theta)$, so the corresponding curvatures of the two branches coincide at each pole. The strict result excludes the extremal limit $c=0$, where the factorization instead gives identical intrinsic metrics. The singular critical quotient $\lambda=1$ and the degenerate branch map $\lambda=0$ are also excluded.

The common area form is $\rho_A^2\sin\theta\,d\theta\,d\widehat\varphi$, so $u=(1+\cos\theta)/2$ is the same area fraction on both horizons. Their latitude circumference is $L_\varphi=2\pi\rho_A^2\sqrt f$, whereas their meridional line element is $\sqrt{\mathcal X}\,d\theta$. Thus every nonpolar latitude is strictly shorter on the supercritical branch, while every corresponding meridional segment is longer. In particular, the pole-to-pole meridian distance is larger and the maximum axial-orbit circumference is smaller.

These geometric differences also exclude an isometry
of the underlying unmarked two-surfaces. Both horizon
metrics are smooth, real-analytic rotational metrics
on $S^2$, with an effective axial $U(1)$ action whose
orbits have two fixed points at the poles.

The isometry group of a compact Riemannian manifold
is compact. For a rotational two-sphere, its connected
isometry group contains the axial $U(1)$ action.
If the metric is nonround, this axial symmetry is the
only continuous isometry. Indeed, any additional
independent Killing field would enlarge the connected
isometry group to a group of dimension greater than
one. The classification of connected compact group
actions on $S^2$ then requires the full $SO(3)$
symmetry, which makes the metric round. Hence, for
a nonround rotational two-sphere, the axial $U(1)$
action is unique.

If either of the two metrics were round, an isometry
would require the other to be round as well. Since
the two horizons have the same area, their normalized
latitude-circumference profiles would then coincide,
contrary to the strict inequality (3.10). Thus neither
metric can be round under an assumed isometry, and
any isometry between them must conjugate their unique
connected axial $U(1)$ actions.

Consequently, such an isometry can only preserve or
reverse the axial orbit coordinate, corresponding
to $u\mapsto u$ or $u\mapsto1-u$. The maximum
latitude-orbit circumference is invariant under
isometry, but the strict inequality (3.10) gives
\[
\max_{u\in[0,1]}L_{\phi,\mathrm{super}}(u)
<
\max_{u\in[0,1]}L_{\phi,\mathrm{sub}}(u).
\]
This contradiction proves that the two horizon
metrics are not isometric, even when the axial
marking is discarded.

\subsection{Strict splitting of the axisymmetric horizon spectrum}
\label{sm:ad:axisymmetric-spectrum}

The strict shape inequality separates every positive axisymmetric scalar eigenvalue of the intrinsic horizon Laplacian. The comparison uses the same regular nonextremal pair and common area radius $\rho_A$ as above; it includes balanced external fields.

For $b\in\{{\rm sub},{\rm super}\}$, write $f_b(u)=f_b(\theta(u))$ in the existing area coordinate $u=(1+\cos\theta)/2$. The axial reduction follows the standard formulation~\cite{AbreuFreitas2002}:
\begin{align}
h_b&=\frac{4}{f_b}\,du^2+\rho_A^4f_b\,d\widehat\varphi^2,\notag\\
\mathscr L_bv&\equiv(-\Delta_{h_b})v
=-\frac14\frac{d}{du}\left(f_b\frac{dv}{du}\right),
\quad \partial_{\widehat\varphi}v=0.
\label{sm:ad:spectral-operator}
\end{align}
After azimuthal integration both area measures are $4\pi\rho_A^2\,du$, so the same $L^2(0,1;du)$ inner product applies. Axis regularity gives
\begin{equation}
f_b=\begin{cases}
4u/\rho_A^2+O(u^2),&u\to0,\\
4(1-u)/\rho_A^2+O((1-u)^2),&u\to1.
\end{cases}
\label{sm:ad:spectral-poles}
\end{equation}
Thus both profiles are uniformly comparable to $u(1-u)$ for each fixed regular pair. The smooth horizon sphere induces the nonnegative self-adjoint realization of $\mathscr L_b$. Equivalently, it is the Friedrichs operator of
\begin{align}
\mathfrak q_b[v]&=\frac14\int_0^1f_b|v'|^2\,du,\notag\\
\mathcal V&=\overline{C^\infty([0,1])}^{\|\cdot\|_{\mathcal V}},\notag\\
\|v\|_{\mathcal V}^2&=\int_0^1\left[|v|^2+u(1-u)|v'|^2\right]du.
\label{sm:ad:spectral-form}
\end{align}
This common form domain has compact embedding in $L^2(0,1)$, by its equivalent axisymmetric $H^1$ norm on the round sphere. Smooth eigenfunctions satisfy $f_bv'\to0$ at both poles; no Dirichlet condition $v=0$ is imposed there. Logarithmic endpoint solutions have infinite form energy and are excluded.

\emph{Spectral splitting theorem.} Let $\Lambda_{n,b}^{(0)}$ be the ordered axisymmetric eigenvalues of $-\Delta_{h_b}$, counted with multiplicity from $n=0$. For every pair in the strict-geometry theorem,
\begin{align}
\Lambda_{0,{\rm super}}^{(0)}&=\Lambda_{0,{\rm sub}}^{(0)}=0,\notag\\
0<\Lambda_{n,{\rm super}}^{(0)}&<\Lambda_{n,{\rm sub}}^{(0)},\qquad n\ge1.
\label{sm:ad:spectral-splitting}
\end{align}
The dimensionless eigenvalues $\rho_A^2\Lambda_{n,b}^{(0)}$ obey the same ordering because $\rho_A$ is common.

Proof: The zero eigenspace consists of constants on each connected sphere. Every nonzero function in the common mean-zero space $\mathcal V_0=\{v\in\mathcal V:\int_0^1v\,du=0\}$ is nonconstant. Hence the strict profile inequality gives
\begin{align}
&\mathfrak q_{\rm sub}[v]-\mathfrak q_{\rm super}[v]\notag\\
&\qquad=\frac14\int_0^1(f_{\rm sub}-f_{\rm super})|v'|^2\,du>0.
\label{sm:ad:spectral-form-gap}
\end{align}
For $n\ge1$, let $V_n\subset\mathcal V_0$ be spanned by the first $n$ positive subcritical eigenfunctions. On its compact unit sphere, the continuous positive form difference has a minimum $\delta_n>0$. The min--max principle then yields
\begin{align}
\Lambda_{n,{\rm super}}^{(0)}
&\le\max_{\substack{v\in V_n\\\|v\|_2=1}}\mathfrak q_{\rm super}[v]\notag\\
&\le\Lambda_{n,{\rm sub}}^{(0)}-\delta_n
<\Lambda_{n,{\rm sub}}^{(0)}.
\label{sm:ad:spectral-minmax}
\end{align}
This proves the strict ordering at every positive index; it does not require a gap uniform in $n$ or in the solution parameters.

In the regular extremal limit $c=0$, the intrinsic metrics coincide and so do their spectra. The singular critical quotient and degenerate branch map remain excluded. For a nonzero azimuthal integer $q$, separation with $e^{iq\widehat\varphi}$ instead gives
\begin{equation}
\mathscr L_{b,q}v=-\frac14(f_bv')'
+\frac{q^2}{\rho_A^4f_b}v,\qquad q\ne0.
\label{sm:ad:spectral-nonaxisymmetric}
\end{equation}
The two terms contribute oppositely to the spectral
comparison, yielding a strict ordering for the
axisymmetric spectrum. We cannot infer from this
result how the non-axisymmetric spectrum is ordered
or whether either black-hole branch is dynamically
stable.

\section{Exact reference backgrounds and uniform angular matching}
\label{sm:eh:m:sec:background}

We take $m>|a|$, ${E_0B_0}\ne0$ and $C>0$. We first extract the two polar backgrounds from the black-hole solution, then verify the extracted fields independently as exact solutions of the full EMD equations.

\subsection{Step 1: Extraction from the two polar regions}
\label{sm:eh:m:sec:extract}

Introduce Weyl coordinates $\varpi=\sqrt{\Delta}\sin\theta$ and $z=(\rho-m)\cos\theta$, and put $\Xi=\Delta+c^2\sin^2\theta$. The exact black-hole metric~\mainEq{eq:metric}{2.2} can be rewritten as
\begin{align}
&ds^2=H\left[-q^{-1}dt^2+l(d\varpi^2+dz^2)\right] +\frac{q\varpi^2}{H}(d\varphi-\omega dt)^2,\notag\\
&q=\frac{s}{\varpi^2},\quad l=\frac{\Sigma}{\Xi}.
\label{sm:eh:m:eq:weyl-extraction}
\end{align}
The ratio $q$ has a smooth axis limit. For the north and south polar regions, respectively, set $z=\sigma L$, with $\sigma=+1$ and $-1$, and take $L\to\infty$ with $0\le\varpi\le\varpi_*$ for fixed $\varpi_*$. The inverse coordinate relations give
\begin{align}
&\rho=m+L+\frac{\varpi^2}{2L}+O(L^{-3}),\notag\\
&\cos\theta=\sigma\left[1-\frac{\varpi^2}{2L^2}+O(L^{-4})\right].
\label{sm:eh:m:eq:polar-tube-limit}
\end{align}
Substituting these relations into the exact functions~\mainEq{eq:metric-functions}{2.3} and~\mainEq{eq:app-currents}{2.6} yields
\begin{align}
&s=\varpi^2\left[1+\frac{2m}{L}+O(L^{-2})\right],\quad l=1+O(L^{-1}),\notag\\
&\mathcal{W}=2\sigma ma+O(L^{-4}),\quad \frac{2ma\rho}{\mathscr A}=O(L^{-3}),\notag\\
&h_K=-2z+2m\sigma+O(L^{-2}),\notag\\
&K_K=-16\sigma ma z+4m^2a+O(L^{-2}).
\label{sm:eh:m:eq:polar-field-extraction}
\end{align}
Thus $q,l\to1$; the background factors and matter potentials are obtained by evaluating Eqs.~\mainEq{eq:field-factors}{2.4} and~\mainEq{eq:fields}{2.5} at $s=\varpi^2$, $\mathcal{W}=2\sigma ma$. The functions $\omega$ and $V$ generally grow linearly with $z$; their asymptotic affine parts must retain both the linear and constant terms. Equation~\mainEq{eq:fields}{2.5} fixes
\begin{align}
&\mathcal{W}_\sigma=2\sigma ma,\quad \omega_0=m^2aE_0^2B_0^2,\notag\\
&V_{0,\sigma}=E_0m(\sigma+\zeta/2),\quad h_{b,\sigma}=-2z+2m\sigma,\notag\\
&K_{b,\sigma}=-16\sigma ma z+4m^2a,\quad \sigma=\pm1.
\label{sm:eh:m:eq:map}
\end{align}
The pole label $\sigma$ is retained on every reference
field. With the extracted constants in
Eq.~\eqref{sm:eh:m:eq:map}, their explicit radial
functions are
\begin{align}
&H_{b,\sigma,+}
=(1+E_0B_0\mathcal{W}_\sigma)^2
+E_0^2\varpi^2
+\frac{E_0^2B_0^2\varpi^4}{4},\notag\\
&H_{b,\sigma,-}
=(1-E_0B_0\mathcal{W}_\sigma)^2
+B_0^2\varpi^2
+\frac{E_0^2B_0^2\varpi^4}{4},\notag\\
&H_{b,\sigma}
=\sqrt{H_{b,\sigma,+}H_{b,\sigma,-}},\notag\\
&\psi_{b,\sigma}
=\frac{\begin{gathered}
-2E_0\mathcal{W}_\sigma(1-E_0B_0\mathcal{W}_\sigma)
\\+B_0\varpi^2
+(E_0^2B_0/2)\varpi^4
\end{gathered}}{
2H_{b,\sigma,-}
}.
\label{sm:eh:m:eq:background-factors}
\end{align}

A prime denotes $d/d\varpi$. The corresponding
metric and matter fields are
\begin{align}
&ds_{b,\sigma}^2={}
H_{b,\sigma}(-dt^2+d\varpi^2+dz^2)
+\frac{\varpi^2}{H_{b,\sigma}}
(d\varphi-\omega_{b,\sigma}dt)^2,\notag\\
&\phi_{b,\sigma}={}
\frac{\sqrt3}{4}
\log\frac{H_{b,\sigma,+}}{H_{b,\sigma,-}},\notag\\
&A_{b,\sigma}={}
V_{b,\sigma}dt
+\psi_{b,\sigma}
(d\varphi-\omega_{b,\sigma}dt),\notag\\
&\omega_{b,\sigma}={}
-2E_0^2B_0^2\mathcal{W}_\sigma z+\omega_0,\notag\\
&V_{b,\sigma}={}
-E_0(1+2E_0B_0\mathcal{W}_\sigma)z+V_{0,\sigma},\notag\\
&F_{b,\sigma}={}
\left[
E_0(1+2E_0B_0\mathcal{W}_\sigma)
-2E_0^2B_0^2\mathcal{W}_\sigma\psi_{b,\sigma}
\right]
dt\wedge dz\notag\\
&+\psi_{b,\sigma}'\,d\varpi
\wedge(d\varphi-\omega_{b,\sigma}dt).
\label{sm:eh:m:eq:background}
\end{align}

The common constant $\omega_0$ fixes the angular frame.
In each bounded polar tube, the full solution approaches
the corresponding reference fields with
\begin{equation}
\omega-\omega_{b,\sigma}=O(L^{-2}),\quad V-V_{b,\sigma}=O(L^{-2}).
\label{sm:eh:m:eq:polar-potential-matching}
\end{equation}
The two reference backgrounds are extracted from the
polar limits of the exact black-hole solution.

\subsection{Step 2: Exactness of the two polar reference backgrounds}
\label{sm:eh:m:sec:verify}

For the independent verification, first regard the constant twist value $\mathcal{W}_\sigma$ and the constants $\omega_0,V_{0,\sigma}$ as arbitrary in Eq.~\eqref{sm:eh:m:eq:background}, before imposing their extracted values in Eq.~\eqref{sm:eh:m:eq:map}. Substituting these four-dimensional fields directly into Eq.~\eqref{sm:ad:emd-equations} at finite $(\varpi,z)$ verifies identities for a general constant twist, and therefore for both selected polar values. Set
\begin{align}
&k_{b,\sigma}=2E_0^2B_0^2\mathcal{W}_\sigma,\notag\\
&j_{b,\sigma}=E_0(1+2E_0B_0\mathcal{W}_\sigma)-k_{b,\sigma}\psi_{b,\sigma},\notag\\
&Z_{b,\sigma}=e^{-2\sqrt3\phi_{b,\sigma}}=\left(\frac{H_{b,\sigma,-}}{H_{b,\sigma,+}}\right)^{3/2}.
\label{sm:eh:m:eq:reference-4d-definitions}
\end{align}
In the orthonormal coframe
\begin{align}
&e^{\hat0}=\sqrt{H_{b,\sigma}}\,dt,\quad e^{\hat1}=\sqrt{H_{b,\sigma}}\,d\varpi,\notag\\
&e^{\hat2}=\sqrt{H_{b,\sigma}}\,dz,\quad e^{\hat3}=\frac{\varpi}{\sqrt{H_{b,\sigma}}}(d\varphi-\omega_{b,\sigma}dt),
\label{sm:eh:m:eq:reference-4d-frame}
\end{align}
the field strength is $F_{b,\sigma}=(j_{b,\sigma}/H_{b,\sigma})e^{\hat0}\wedge e^{\hat2}+(\psi_{b,\sigma}'/\varpi)e^{\hat1}\wedge e^{\hat3}$. Define the weighted squared components
\begin{equation}
\mathcal E_{b,\sigma}=\frac{H_{b,\sigma,-}}{H_{b,\sigma,+}^{\,2}}j_{b,\sigma}^2,\quad \mathcal B_{b,\sigma}=\frac{H_{b,\sigma,-}^{\,2}}{H_{b,\sigma,+}}\left(\frac{\psi_{b,\sigma}'}{\varpi}\right)^2.
\label{sm:eh:m:eq:reference-4d-matter}
\end{equation}
Direct calculation of the four-dimensional Christoffel symbols and Ricci tensor gives zero off-diagonal frame components and
\begin{align}
&H_{b,\sigma} R_{\hat0\hat0}
 =H_{b,\sigma} R_{\hat3\hat3}=-H_{b,\sigma} R_{\hat2\hat2}\notag\\
&=\frac12\left[(\log H_{b,\sigma})''+\frac{(\log H_{b,\sigma})'}{\varpi}
 -\frac{k_{b,\sigma}^2\varpi^2}{H_{b,\sigma}^2}\right]\notag\\
&=\mathcal E_{b,\sigma}+\mathcal B_{b,\sigma},\notag\\
&H_{b,\sigma} R_{\hat1\hat1}
 =-\frac{H_{b,\sigma}''}{2H_{b,\sigma}}+\frac{3H_{b,\sigma}'}{2\varpi H_{b,\sigma}}\notag\\
&=2(\phi_{b,\sigma}')^2+\mathcal E_{b,\sigma}+\mathcal B_{b,\sigma}.
\label{sm:eh:m:eq:reference-4d-einstein}
\end{align}
The final equalities follow by inserting Eq.~\eqref{sm:eh:m:eq:background-factors}; their numerators vanish as polynomial identities for arbitrary $E_0,B_0,\mathcal{W}_\sigma$. These are precisely all ten four-dimensional Einstein equations with the scalar and Maxwell stress tensors in Eq.~\eqref{sm:ad:emd-equations}.

For the Maxwell equation, $\sqrt{-g_{b,\sigma}}=H_{b,\sigma}\varpi$. Directly raising both indices of $F_{b,\sigma}$ makes the $t,\varpi,z$ components of $\partial_\mu(\sqrt{-g_{b,\sigma}}Z_{b,\sigma}F_{b,\sigma}^{\mu\nu})$ vanish identically; its remaining component is
\begin{equation}
\partial_\varpi\!\left[\frac{H_{b,\sigma,-}^{\,2}}{H_{b,\sigma,+}}
          \frac{\psi_{b,\sigma}'}{\varpi}\right]
 -k_{b,\sigma}\varpi\frac{H_{b,\sigma,-}}{H_{b,\sigma,+}^{\,2}}j_{b,\sigma}=0.
\label{sm:eh:m:eq:reference-4d-maxwell}
\end{equation}
Similarly, the direct four-dimensional contractions give $\Box_{b,\sigma}\phi_{b,\sigma}=(\phi_{b,\sigma}''+\phi_{b,\sigma}'/\varpi)/H_{b,\sigma}$ and $Z_{b,\sigma}F_{b,\sigma}^2=2(\mathcal B_{b,\sigma}-\mathcal E_{b,\sigma})/H_{b,\sigma}$. Hence the dilaton equation is the identity
\begin{equation}
\phi_{b,\sigma}''+\frac{\phi_{b,\sigma}'}{\varpi}
 +\sqrt3(\mathcal B_{b,\sigma}-\mathcal E_{b,\sigma})=0.
\label{sm:eh:m:eq:reference-4d-dilaton}
\end{equation}
Equations~\eqref{sm:eh:m:eq:reference-4d-maxwell} and~\eqref{sm:eh:m:eq:reference-4d-dilaton} likewise have identically zero polynomial numerators. Finally, $F_{b,\sigma}=dA_{b,\sigma}$ verifies all components of $dF_{b,\sigma}=0$. This establishes the full four-dimensional EMD system throughout every regular patch with $\varpi>0$ and $z\in\mathbb R$. The field equations are independent of the frame and
gauge constants $\omega_0$ and $V_{0,\sigma}$.

At the axis, $H_{b,\sigma}(0)=|1-E_0^2B_0^2\mathcal{W}_\sigma^2|$. For a positive value, smoothness fixes the angular period to $2\pi|1-E_0^2B_0^2\mathcal{W}_\sigma^2|$; the even radial functions and a local Maxwell gauge patch extend the fields smoothly to $\varpi=0$. With the extracted value $\mathcal{W}_\sigma=2\sigma ma$, both backgrounds therefore have the common period $2\pi C$ in Eq.~\eqref{sm:ad:common-period} and the horizon's angular normalization. Thus, each polar reference background is independently an exact four-dimensional EMD solution, rather than merely a leading-order asymptotic approximation to the black-hole metric.

\subsection{Uniform asymptotic matching of the full EMD fields}

\label{sm:eh:m:sec:matching}
This section proves that the full rotating black-hole solution with dual external fields not only approaches the northern and southern cylindrical reference backgrounds at fixed angles, but also that the Maxwell electromagnetic field and its first derivatives converge uniformly to the corresponding backgrounds over each hemisphere.

For each sign, use the reference~\eqref{sm:eh:m:eq:background} with~\eqref{sm:eh:m:eq:map} on the closed hemisphere $\sigma \cos\theta \ge0$. Extend each comparison into a fixed cone away from the opposite axis to obtain an overlap. Use $q,l$ from Eq.~\eqref{sm:eh:m:eq:weyl-extraction} and, for each fixed $\sigma$, the orthonormal coframe in Eq.~\eqref{sm:eh:m:eq:reference-4d-frame}, omitting hats on its frame indices. Define

\begin{equation}
b=\frac{H}{H_{b,\sigma}},\quad \beta_\sigma=\frac{\varpi(\omega-\omega_{b,\sigma})}{H_{b,\sigma}}.
\label{sm:eh:m:eq:coframe}
\end{equation}

Write $G_{ab}$ and $F_{ab}$ for full-solution components in this reference coframe; the reference metric has components $\eta_{ab}$. Let $\delta G=G-\eta$, $\delta\phi=\phi-\phi_{b,\sigma}$ and $\delta F=F-F_{b,\sigma}$ denote the component differences, and put $\delta\psi=\psi-\psi_{b,\sigma}$. With a maximum component norm, define

\begin{align}
&\mathcal N_\sigma={}|\delta G|+|\delta\phi|+m|\delta F|\notag\\
&+\frac m{\sqrt{H_{b,\sigma}}}\sum_{i=\varpi,z} \left(|\partial_i\delta G|+|\partial_i\delta\phi|+m|\partial_i\delta F|\right).
\label{sm:eh:m:eq:norm}
\end{align}

Matching statement: For every fixed regular parameter point with $m>|a|$, $E_0B_0\ne0$, and $C>0$, there are constants $\rho _0$ and $K$ such that

\begin{align}
&\sup_{\sigma \cos\theta \ge0}\mathcal N_\sigma\le K\frac m\rho ,\quad \sup_{\sigma \cos\theta \ge0}|\beta_\sigma|\le K\frac{m^2}{\rho ^2},\notag\\
&\rho \ge \rho _0.
\label{sm:eh:m:eq:theorem}
\end{align}

We evaluate the fields and their derivatives on the
axes using regular polar Cartesian coordinates.
The derivative norm is defined in
\eqref{sm:eh:m:eq:norm} and includes derivatives along
both orbit-space directions. The constants in the
estimates depend on the fixed parameters and are not
uniform in the limits $C\to0$, $E_0B_0\to0$, or
$c\to0$. We now derive the required estimates.
Throughout the proof, all lengths are measured in
units of $m$, so that the polynomial weights are
dimensionless.

\subsubsection{Exact remainder polynomials}
To obtain uniform estimates up to both symmetry axes,
we retain the exact rational remainders of the Kerr
currents rather than relying only on fixed-angle
asymptotic expansions
\begin{align}
&N_h={}2a^2m\cos\theta \sin^2\theta \rho ^2\notag\\
&+4a^2m^2\cos\theta (\cos^2\theta -3)\rho\notag\\
&+2a^4m\cos\theta \sin^2\theta,\notag\\
&N_K={}2a^3m\sin^6\theta \rho ^3\notag\\
&+8a^3m^2\cos^2\theta (\cos^2\theta -1)(\cos^2\theta -3)\rho ^2\notag\\
&-2a^3m\left[a^2(\cos^2\theta -1)^3+4m^2\cos^2\theta (\cos^2\theta -3)^2\right]\rho\notag\\
&+4a^5m^2(\cos^2\theta -1)(\cos^4\theta -4\cos^2\theta -1).
\label{sm:eh:m:eq:remainders}
\end{align}
Their polynomial factors provide the additional
vanishing powers near both axes, allowing the
remainder estimates and their first derivatives
to hold uniformly in the shrinking polar regions.

\subsubsection{Seed and current estimates}
\begin{align}
&q=1+\frac{2m\rho(\rho^2+a^2)}{\Sigma\Delta},\quad l=1+\frac{2m\rho-m^2\sin^2\theta}{\Xi},\notag\\
&\begin{pmatrix} \rho_\varpi & \theta_\varpi\\ \rho_z & \theta_z \end{pmatrix} = \frac{1}{\Xi} \begin{pmatrix} \varpi(\rho-m) & \sqrt{\Delta}\cos\theta\\ \Delta\cos\theta & -(\rho-m)\sin\theta \end{pmatrix},\notag\\
&\mathcal{W}-\mathcal{W}_\sigma=-\sigma ma(1-\sigma\cos\theta)^2(2+\sigma\cos\theta)\notag\\
&+\frac{ma^3\cos\theta\sin^4\theta}{\Sigma}.
\label{sm:eh:m:eq:exactidentities}
\end{align}

These exact identities give $q-1,l-1=O(\rho ^{-1})$ and their Weyl first derivatives $O(\rho ^{-2})$. On the designated hemisphere, $1-\sigma \cos\theta =\sin^2\theta /(1+\sigma \cos\theta)$, so $\mathcal{W}-\mathcal{W}_\sigma=O(\sin^4\theta)$. Polynomial division of the explicit currents gives

\begin{align}
&h_K=-2\rho \cos\theta +4m\cos\theta +\frac{N_h}{\mathscr A},\notag\\
&K_K=-2ma\rho (3+6\cos^2\theta -\cos^4\theta)\notag\\
&+4m^2a\cos^2\theta (6-\cos^2\theta)+\frac{N_K}{\mathscr A},\notag\\
&\frac{N_h}{\mathscr A}=O(\sin^2\theta /\rho ^2+\rho ^{-3}),\notag\\
&\frac{N_K}{\mathscr A}=O(\sin^6\theta /\rho +\sin^2\theta /\rho ^2+\rho ^{-3}),\notag\\
&3+6\cos^2\theta -\cos^4\theta -8\sigma \cos\theta\notag\\
&=(1-\sigma \cos\theta)^3(3+\sigma \cos\theta).
\label{sm:eh:m:eq:division}
\end{align}
\begin{align}
&\delta h=h_K-h_{b,\sigma}=O(\sin^2\theta +\rho ^{-3}),\notag\\
&\delta K=K_K-K_{b,\sigma}=O(\rho \sin^6\theta +\sin^2\theta +\rho ^{-3}).
\label{sm:eh:m:eq:currentbounds}
\end{align}

The explicit remainder polynomials are given in Eq.~\eqref{sm:eh:m:eq:remainders}. For large $\rho $, $\mathscr A\ge \rho ^4$, while $\Sigma,\Delta,\Xi$ have positive lower bounds proportional to $\rho ^2$. Differentiating these explicit rational expressions with~\eqref{sm:eh:m:eq:exactidentities} supplies the first and second derivative bounds used below.

\subsubsection{Metric and dilaton}

At fixed regular parameters, all reference factors are comparable to $1+\varpi^4$. Since $s-\varpi^2=O(\varpi^2/\rho )$ and $\mathcal{W}-\mathcal{W}_\sigma=O(\varpi^4/\rho ^4)$, their full-solution counterparts have the same positive lower bounds at sufficiently large $\rho $. In particular,

\begin{align}
&k(1+\varpi^4)\le H_{b,\sigma,\pm},H_{b,\sigma}\le K(1+\varpi^4),\notag\\
&\frac{H_\pm}{H_{b,\sigma,\pm}}-1=O(\rho ^{-1}),\quad b-1=O(\rho ^{-1}),\notag\\
&\phi-\phi_{b,\sigma}=O(\rho ^{-1}),\notag\\
&\beta_\sigma=\frac\varpi{H_{b,\sigma}}\left(\frac{2ma\rho}{\mathscr A}+\frac{{E_0^{2}B_0^{2}}}{4}\delta K\right)=O(\rho ^{-2}).
\label{sm:eh:m:eq:factorbounds}
\end{align}

The last estimate uses $\sin^2\theta =\varpi^2/\Delta$, $\varpi\le\sqrt\Delta=O(\rho )$ and the boundedness of $\varpi^3/(1+\varpi^4)$. Thus the coordinate shift can grow with distance while its physical-frame mismatch decays. The exact metric components are

\begin{align}
&G_{00}=-\frac bq+\frac qb\beta_\sigma^2,\quad G_{03}=-\frac qb\beta_\sigma,\quad G_{33}=\frac qb,\notag\\
&G_{11}=G_{22}=bl.
\label{sm:eh:m:eq:frameG}
\end{align}

Equations~\eqref{sm:eh:m:eq:factorbounds}--\eqref{sm:eh:m:eq:frameG} yield $\delta G=O(\rho ^{-1})$. Their rational derivatives give $H_{b,\sigma}^{-1/2}\partial_i\delta G=O(\rho ^{-1})$ and $H_{b,\sigma}^{-1/2}\partial_i\delta\phi=O(\rho ^{-1})$. The transverse derivative need not gain an additional power of $\rho $ in a fixed cylindrical tube; this is why the derivative weight in~\eqref{sm:eh:m:eq:norm} uses physical distance, rather than an extra isotropic power of $\rho $.

\subsubsection{Maxwell field and first derivatives}

\begin{align}
&F=(dV-\psi d\omega)\wedge dt+d\psi\wedge(d\varphi-\omega dt),\notag\\
&\mathsf E_i=\frac{V_i-\psi\omega_i-\psi_i\delta\omega}{H_{b,\sigma}},\quad \mathsf B_\varpi=\frac{\psi_\varpi}{\varpi},\quad \mathsf B_z=\frac{\psi_z}{\varpi},\notag\\
&\delta\omega=\omega-\omega_{b,\sigma}.
\label{sm:eh:m:eq:frameF}
\end{align}

Here $\mathsf E_i$ multiplies $e^i\wedge e^0$ and $\mathsf B_i$ multiplies $e^i\wedge e^3$, for $i=\varpi,z$, with $e^\varpi\equiv e^1$ and $e^z\equiv e^2$. The key cancellation follows directly from an exact identity among the potentials:

\begin{align}
&E_0-{E_0B_0}\psi=\frac{E_0[(1-{E_0B_0} \mathcal{W})+B_0^2s/2]}{H_-},\notag\\
&\psi=\frac1{B_0}-\frac{(1-{E_0B_0} \mathcal{W})+B_0^2s/2}{B_0H_-}.
\label{sm:eh:m:eq:cancellation}
\end{align}

Along the interpolation from $(\varpi^2,\mathcal{W}_\sigma)$ to $(s,\mathcal{W})$, positive denominator bounds and polynomial division imply

\begin{align}
&|E_0-{E_0B_0}\psi|\le\frac K{1+s},\notag\\
&|\partial_s^k\psi|\le\frac{K_k}{(1+s)^{k+1}}\quad(k\ge1),\notag\\
&|\partial_{\mathcal{W}}\partial_s^k\psi|\le\frac{K_k}{(1+s)^{k+2}}\quad(k\ge0),\notag\\
&\delta\psi=O\left(\frac{\varpi^2}{\rho (1+\varpi^2)^2} +\frac{\sin^4\theta}{(1+\varpi^2)^2}\right)\notag\\
&=O\left(\frac{\varpi^2}{\rho (1+\varpi^2)^2}\right).
\label{sm:eh:m:eq:psibounds}
\end{align}

Only finitely many derivatives, through total order three, are needed. On bounded $s$ intervals they are bounded by regularity; at large $s$ their powers follow from~\eqref{sm:eh:m:eq:cancellation}. Higher $\mathcal{W}$ derivatives are rational and obey stronger large-$s$ decay. Applying the inverse Jacobian in~\eqref{sm:eh:m:eq:exactidentities} to $\delta s$ and $\delta \mathcal{W}$ gives

\begin{align}
&\frac{\delta\psi_\varpi}{\varpi}=O(\rho ^{-1}),\quad \frac{\psi_z}{\varpi}=O(\rho ^{-1}),\notag\\
&H_{b,\sigma}^{-1/2}\partial_i\left(\frac{\delta\psi_\varpi}{\varpi},\frac{\psi_z}{\varpi}\right)=O(\rho ^{-1}).
\label{sm:eh:m:eq:Bbound}
\end{align}

The electric current can be written before subtraction as

\begin{align}
&dV-\psi d\omega=\frac{E_0}{2}dh_K\notag\\
&+\frac{E_0B_0}4(E_0-{E_0B_0}\psi)dK_K-\psi d\!\left(\frac{2ma\rho}{\mathscr A}\right),\notag\\
&\delta(dV-\psi d\omega)=\frac{E_0}{2}d\delta h\notag\\
&+\frac{E_0B_0}4\Bigl[(E_0-{E_0B_0}\psi)d\delta K\notag\\
&-{E_0B_0}\delta\psi\,dK_{b,\sigma}\Bigr]
-\psi d\!\Bigl(\frac{2ma\rho}{\mathscr A}\Bigr).
\label{sm:eh:m:eq:electricdifference}
\end{align}

Add the reference-frame correction $-d\psi\,\delta\omega$ and divide by $H_{b,\sigma}$. The bounds~\eqref{sm:eh:m:eq:currentbounds}, \eqref{sm:eh:m:eq:psibounds} and the inverse Jacobian give an $O(\rho ^{-1})$ difference. One further $\varpi$ or $z$ derivative uses at most second derivatives of the explicit currents and third derivatives of the rational potential. The same positive denominators control them, giving the Maxwell part of~\eqref{sm:eh:m:eq:theorem}.

The factors $1/\varpi$ in~\eqref{sm:eh:m:eq:frameF} do not produce axis singularities. The scalar potentials are even analytic functions of $\varpi$ near either exterior axis; $\delta\psi$ vanishes on that axis and its axis value is independent of $z$. Also $G_{11}=G_{33}$ at the axis and $\beta_\sigma$ is divisible by $\varpi$. Passing to smooth Cartesian frames therefore preserves the same bounds and their first derivatives.

\subsubsection{Overlap of the two reference descriptions}

On a fixed cone $|\cos\theta |\le1-\delta$, with $0<\delta<1$, one has $\varpi\asymp \rho $. The coordinate transition is the identity in $(t,\varpi,z,\varphi)$, with the common period~\eqref{sm:ad:common-period}. The relative differences $H_{b,+,\pm}/H_{b,-,\pm}-1$ and the dilaton difference $\phi_{b,+}-\phi_{b,-}$ are $O(\rho ^{-4})$. The coframe transition has shear

\begin{equation}
\frac{\varpi(\omega_{b,+}-\omega_{b,-})}{H_{b,+}}
=\frac{-8ma{E_0^{2}B_0^{2}}\varpi z}{H_{b,+}}=O(\rho ^{-2}).
\label{sm:eh:m:eq:overlap}
\end{equation}

The rational Maxwell expressions and their first physical derivatives also have vanishing differences there. Thus the two exact reference models give compatible asymptotic data on the overlap, and~\eqref{sm:eh:m:eq:theorem} covers the full angular range. 

Together with the regularity of the exterior established
in Sec.~\ref{sm:ad:C}, these results connect the rotating
Killing horizon to the two asymptotic cylindrical
backgrounds. The spatial end lies at infinite proper
distance. Indeed, on a constant-$t$ slice, the orbit-space
metric is
\[
\frac{H\Sigma}{\Xi}
(d\varpi^2+dz^2),
\]
whose coefficient is bounded below by a positive
constant outside a sufficiently large sphere.
Consequently, every spatial curve extending to infinity
has infinite length. The global event-horizon
identification is established separately through the
causal structure of the selected future-ingoing
extension.

\section{Global hyperbolicity of the open exterior}

\label{sm:eh:c:sec:causal}
We prove global hyperbolicity by showing that the
constant-$t$ hypersurfaces are Cauchy surfaces
\cite{Geroch1970,BernalSanchez2005}. The proof relies
on the fact that $t$ is a global time function and
that the radial causal cones prevent causal curves
from reaching either radial end within a finite
time interval.
\subsection{Cauchy time for the complete open exterior}

For $\rho >\rho _+$, direct inversion gives $g^{tt}<0$. Every future causal tangent obeys the following inequality after the nonnegative angular square is removed:

\begin{equation}
\begin{aligned}
&g^{tt}=-\frac{\mathscr A}{H\Sigma\Delta}<0,\\
&-\frac{\Delta}{\mathscr A}dt^2+\frac{d\rho ^2}{\Delta}+d\theta^2\le0,\\
&dt\ge\frac{\sqrt{\mathscr A}}\Delta|d\rho |.
\end{aligned}
\label{sm:eh:c:eq:cone}
\end{equation}

Conversely, every projected causal curve has a causal lift obtained from $d\varphi=\omega dt$. Thus the projected cone is exactly the Kerr orbit-space cone, independent of $E_0,B_0$. This does not identify the full four-dimensional causal spaces, because the angular quotient and the extension still matter.

In the exterior $\mathscr A\ge \rho ^4$. The monotone optical coordinate

\begin{equation}
\begin{aligned}
&\mathfrak r(\rho )=\int^\rho \frac{u^2}{u^2-2mu+a^2}\,du,
\quad |d\mathfrak r|\le dt,\\
&\mathfrak r(\rho _+^+)=-\infty,\quad \mathfrak r(\infty)=+\infty.
\end{aligned}
\label{sm:eh:c:eq:optical}
\end{equation}

Suppose that $t$ has a finite upper bound $T$ on a future-inextendible causal curve. Starting at $t_0<T$, Eq.~\eqref{sm:eh:c:eq:optical} bounds the total radial optical displacement by $T-t_0$. Since $\mathfrak r$ tends to opposite infinities at the two radial ends, the entire tail stays in $[t_0,T]\times[\rho_1,\rho_2]\times S^2$, with $\rho_+<\rho_1\le\rho_2<\infty$. Both regular axes belong to this compact set.

To see explicitly why such total imprisonment is impossible, choose an auxiliary smooth Riemannian metric on the exterior. Over the compact set, the future causal vectors of unit Riemannian length form a compact set on which $dt$ has a strictly positive minimum, since $\nabla t$ is timelike. The Riemannian length of the tail is therefore bounded by a constant times $T-t_0$. The curve has a future endpoint in the compact set and extends from it along a timelike segment, contradicting future inextendibility. The past argument is identical. Thus $t$ ranges over $\mathbb R$ on every inextendible causal curve and is strictly increasing. Each level is a Cauchy surface with topology $(\rho_+,\infty)\times S^2$.

This proof applies in the supercritical $\zeta$ ranges as well. A growing ergoregion does not invalidate it: a stationary Killing vector can become spacelike while the gradient of a time function remains timelike.

\subsection{The axial future escape end and the global event horizon}
\label{sm:global-event-horizon}

We identify the future outer event horizon relative to
a family of outgoing null geodesics on the regular north
axis. The construction uses the causal accessibility
of this family from the exterior and the one-way causal
structure of the selected future-ingoing extension.

Throughout this subsection, we assume
\[
m>|a|>0,\qquad E_0B_0\ne0,\qquad C>0,
\]
and consider the single $\rho>0$ sheet of the
future-ingoing extension $\mathcal M_{\rm in}$
containing the original exterior. We write
\[
\begin{gathered}
\mathcal R=\rho^2+a^2,\qquad \rho_\pm=m\pm\sqrt{m^2-a^2},\\
\Delta=(\rho-\rho_+)(\rho-\rho_-).
\end{gathered}
\]
The angular coordinate is normalized by
$\varphi=C\widehat\varphi$, with
$\widehat\varphi$ having period $2\pi$.

\paragraph{1. Outgoing null geodesics on the regular axis.}

On the north axis, the induced two-dimensional metric
takes the form
\begin{equation}
\begin{gathered}
ds_{\rm ax}^2=C f_{\rm ax}(\rho)\left(-dt^2+d\rho_*^2\right),\\
f_{\rm ax}(\rho)=\frac{\Delta}{\mathcal R},\qquad
\frac{d\rho_*}{d\rho}=\frac{1}{f_{\rm ax}}.
\end{gathered}
\label{sm:eh:axial-metric}
\end{equation}
The outgoing radial null curves are
\begin{equation}
\gamma_u:\qquad
u=t-\rho_*=\mathrm{constant},
\qquad u\in\mathbb R.
\label{sm:eh:null-family}
\end{equation}
The regular axis is totally geodesic as the fixed-point
set of the axial $U(1)$ action. Consequently, these
curves are also four-dimensional null geodesics.

Let $\lambda_{\rm aff}$ be an affine parameter.
The conserved Killing energy associated with
$\partial_t$ is
\begin{equation}
C f_{\rm ax}\frac{dt}{d\lambda_{\rm aff}}=E,
\qquad E>0.
\end{equation}
Using the outgoing null condition
$d\rho_*/d\lambda_{\rm aff}=dt/d\lambda_{\rm aff}$,
we obtain
\begin{equation}
\frac{d\rho}{d\lambda_{\rm aff}}
=
\frac{E}{C}.
\label{sm:eh:affine-radius}
\end{equation}
Thus every member of the family reaches
$\rho\rightarrow\infty$ only at infinite affine
parameter.

The asymptotic behavior of these rays can also be
displayed in the conformal geometry induced on the
axis. Introducing
\[
y=\frac{1}{\rho},
\qquad
f(y)=
\frac{1-2my+a^2y^2}{1+a^2y^2},
\]
and using $u=t-\rho_*$, the conformally rescaled
axial metric becomes
\begin{equation}
y^2ds_{\rm ax}^2
=
C\left[-y^2f(y)\,du^2+2\,du\,dy\right].
\label{sm:eh:axial-conformal-metric}
\end{equation}
Its determinant is $-C^2$, and its inverse metric
has
\[
(g_{\rm ax}^{-1})^{yy}
=
\frac{y^2f(y)}{C}.
\]
The rescaled metric extends smoothly and remains
nondegenerate to $y=0$, where the boundary is null.
Each outgoing ray $\gamma_u$ therefore has a
well-defined endpoint at $y=0$, labelled by its
retarded time $u$. These endpoints define the
future escape end along the regular north axis.

\paragraph{2. Every exterior event can reach the outgoing family.}

We first show that every point of the original
exterior can reach the regular north axis along
a future-directed timelike curve.

At fixed $\rho>\rho_+$, choose
\[
d\rho=0,\qquad d\varphi=\omega\,dt.
\]
The metric then reduces to
\[
ds^2
=
H\Sigma
\left(
-\frac{\Delta}{\mathscr A}\,dt^2
+d\theta^2
\right).
\]
Since
\[
\mathscr A
=
\mathcal R^2-a^2\Delta\sin^2\theta
\leq\mathcal R^2,
\]
the choice
\begin{equation}
\frac{d\theta}{dt}
=
-\frac{\sqrt{\Delta}}{2\mathcal R}
\label{sm:eh:axis-reaching-curve}
\end{equation}
makes the curve strictly timelike:
\[
ds^2
=
H\Sigma
\left(
-\frac{\Delta}{\mathscr A}
+\frac{\Delta}{4\mathcal R^2}
\right)dt^2<0.
\]
The angular interval is finite, so the north axis
is reached in finite coordinate time.

Once on the axis, a future-directed timelike curve
can remain at fixed $\rho$ and wait until it meets
a sufficiently late member of the outgoing family
$\{\gamma_u\}$. Therefore every exterior event lies
in the chronological past of at least one member
of this family. Hence
\begin{equation}
\{\rho>\rho_+\}
\subset
\bigcup_{u\in\mathbb R}
I^-_{\mathcal M_{\rm in}}(\gamma_u).
\label{sm:eh:exterior-inclusion}
\end{equation}

\paragraph{3. Time orientation and one-way propagation through the inner band.}

We now use the regular future-ingoing chart
\begin{equation}
dv_{\rm in}
=
dt+\frac{\mathcal R}{\Delta}\,d\rho,
\qquad
d\Phi
=
d\varphi+
\frac{\mathcal R\omega(\rho,0)}{\Delta}\,d\rho.
\label{sm:eh:ingoing-chart}
\end{equation}
The horizon values of $\omega$ are constant on each
simple horizon, so the azimuthal transformation is
regular at both roots.

For
\[
\tau=v_{\rm in}-\rho,
\]
direct inversion of the four-dimensional metric on $\rho>0$ gives
\begin{equation}
H\Sigma\,g^{-1}(d\tau,d\tau)
=-\bigl(\rho^2+2m\rho+a^2\cos^2\theta\bigr)<0,
\label{sm:eh:time-function}
\end{equation}
and
\begin{equation}
H\Sigma\,g^{-1}(d\tau,d\rho)=2m\rho>0.
\label{sm:eh:time-radial-product}
\end{equation}
Thus $\tau$ is a time function and fixes the
future orientation throughout the selected
positive-$\rho$ extension.

In the intermediate band
\[
\rho_-<\rho<\rho_+,
\]
we have
\begin{equation}
g^{-1}(d\rho,d\rho)
=
\frac{\Delta}{H\Sigma}<0.
\label{sm:eh:radial-timelike}
\end{equation}
Since $\nabla\tau$ is past-directed timelike and
$g(\nabla\tau,\nabla\rho)>0$, the vector
$\nabla\rho$ is future-directed timelike in this
band. Hence every nonzero future-directed causal
tangent $X$ satisfies
\begin{equation}
d\rho(X)=g(\nabla\rho,X)<0,
\qquad
\rho_-<\rho<\rho_+.
\label{sm:eh:future-radial-direction}
\end{equation}
No future-directed causal curve can cross this band
outward.

The same conclusion holds on the future outer
horizon. Its null generator is
\[
K_+
=
\partial_t+
\frac{\omega_+}{C}\partial_{\widehat\varphi},
\qquad
\omega_+=\omega|_{\rho_+},
\]
and the horizon identity is
\begin{equation}
g(K_+,X)
=
\frac{H\Sigma}{\mathcal R_+}\,d\rho(X),
\qquad
\mathcal R_+=\rho_+^2+a^2,
\label{sm:eh:horizon-generator}
\end{equation}
for any tangent vector $X$ at the future outer
horizon. Since $K_+$ is future-directed and null,
the causal inequality $g(K_+,X)\leq0$ for every
future-directed causal $X$ implies
\[
d\rho(X)\leq0
\qquad\text{on }\rho=\rho_+.
\]
Thus no future-directed causal curve can leave
the future outer horizon toward the original
exterior. Points with $0<\rho<\rho_-$ cannot
return to the exterior either, since any such
curve would have to cross the intermediate band
outward.

\paragraph{4. Identification of the global event horizon.}

Define the escape region by the causal past of the
entire outgoing family,
\begin{equation}
\mathcal P_{\rm esc}
:=
\bigcup_{u\in\mathbb R}
I^-_{\mathcal M_{\rm in}}(\gamma_u).
\label{sm:eh:escape-region}
\end{equation}
The exterior reachability result
\eqref{sm:eh:exterior-inclusion}, together with
the one-way causal properties
\eqref{sm:eh:future-radial-direction} and
\eqref{sm:eh:horizon-generator}, gives
\begin{equation}
\mathcal P_{\rm esc}
=
\{\rho>\rho_+\}.
\label{sm:eh:causal-past}
\end{equation}
Consequently, the boundary of this escape region is
\begin{equation}
\mathcal H^+_{\rm out}
=
\partial_{\mathcal M_{\rm in}}\mathcal P_{\rm esc}
=
\{\rho=\rho_+\}_{\rm future}.
\label{sm:eh:global-horizon}
\end{equation}
Thus the future outer Killing horizon is also the
global future event horizon relative to the specified
axial future escape end.

The argument applies to both regular branches,
$|\zeta|<1$ and $|\zeta|>1$, provided
$m>|a|>0$, $E_0B_0\ne0$, and $C>0$.
The critical values $\zeta=\pm1$ are excluded
because the common axial period degenerates.

The result is established in the selected positive-$\rho$
future-ingoing extension, with the axial future escape
end specified above. It identifies the outer event
horizon relative to this end, while the global structure
of other future ends remains to be explored.

\end{document}